\documentclass[12pt,preprint]{aastex}

\slugcomment{}

\shorttitle{Variable stars with 2MASS public images}
\shortauthors{Kouzuma and Yamaoka}

\begin{document}

\title{Properties of large-amplitude variable stars \\detected with 2MASS public images}

\author{Shinjirou Kouzuma and Hitoshi Yamaoka}
\affil{Graduate School of Sciences, Kyushu University, Fukuoka 812-8581, Japan}
\email{kouzuma@phys.kyushu-u.ac.jp}

\begin{abstract}
We present a catalogue of variable stars in the near-infrared wavelength 
detected with overlapping regions of the 2MASS public images, 
and discuss their properties. 
The investigated region is in the direction of the Galactic center 
($-30^\circ \lesssim l \lesssim 20^\circ, |b| \lesssim 20^\circ$), 
which covers the entire bulge. 
We have detected 136 variable stars, 
of which 6 are already-known and 
118 are distributed in $|b| \leq 5^\circ$ region. 
Additionally, 84 variable stars have optical counterparts in DSS images. 
The three diagrams (colour-magnitude, light variance and colour-colour diagrams) 
indicate that most of the detected variable stars should be large-amplitude and long-period variables 
such as Mira variables or OH/IR stars. 
The number density distribution of the detected variable stars implies that 
they trace the bar structure of the Galactic bulge. 
\end{abstract}


\section{Introduction}

The variable stars are useful tracers to study properties of the Galactic bulge.
Among variables, the long-period variable stars 
(such as Semi-Regular Variables (SRVs), Miras and OH/IR stars) 
are detectable even in highly obscured region of the bulge 
owing to their high luminosities. 
These variable stars have been often used for investigating the Galactic bulge
:e.g., measurement of the distance to the Galactic centre using Mira variables 
\citep{Glass1995-MNRAS,Catchpole1999-IAUS,Groenewegen2005-AA,Matsunaga2009-pre}, 
a study of population of the bulge \citep{Glass2001-MNRAS,Groenewegen2005-AA}. 
Therefore, a discovery of variable stars leads to provide us with important information 
about both variable star itself and the Galactic bulge. 

Due to a large amount of dust in the bulge, 
past searches for variable stars have been mainly focused on relatively unobscured fields 
such as Baade's Windows
\citep[e.g., ][]{Evans1976-MNRAS,Glass1995-MNRAS}. 
Near-infrared observation is also better for collecting variable stars in the inner bulge, 
since near-infrared light suffers relatively small interstellar absorption. 
\citet{Glass2001-MNRAS} detected large-amplitude variable stars 
in a $24 \times 24$ arcmin$^2$ area of the Galactic centre from K-band survey, 
and obtained $\sim 400$ objects with periods and amplitudes. 
\citet{Schultheis2000-AA} looked for variable stars by comparing only twice DENIS observations, 
and presented two catalogues with the property of their variable star candidates. 
However, past surveys have only investigated small restricted area (up to a few deg$^2$). 

Various imaging surveys have been performed in this two decades. 
In the imaging surveys, overlapping regions are produced for observing the sky without spatial gaps. 
In other words, there are regions observed multiple times in different epochs 
even if an imaging survey was performed only one time. 
Therefore, variable objects are expected to be detected with the overlapping regions. 
As claimed by \citet{Schultheis2000-AA}, 
$\sim 40\%$ of variable stars can be recovered based on only twice-epoch magnitudes. 
Accordingly, the overlapping regions can be useful tools to search for variable stars. 
However, they have not been used in order to search for variable stars. 
When we can establish the search method based on overlapping region, 
it is also possible to search for variable stars using other survey data 
because survey images always contain overlapping regions. 
Furthermore, we can easily collect many variable stars in widely region 
(over a few tens of deg$^2$) with smaller time 
though we can extract less information compared with variable stars 
derived by an ordinary search (e.g., period, amplitude). 

In this paper, we present a catalogue of variable stars obtained from overlapping region of 2MASS public images.  
In Sect. \ref{Data}, we introduce 2MASS observation and data, 
and discuss the photometric accuracy in an overlapping region. 
In Sect. \ref{Data-Analysis}, after introducing the detection procedure and criteria, 
we show a result of search and estimate the interstellar extinctions. 
In Sect. \ref{Probability}, we discuss detection probability based on twice-epoch magnitudes. 
In Sect. \ref{Discussion}, 
we discuss the near-infrared properties of the detected variable stars and the spatial distribution in the bulge.


\section{Data}\label{Data}

\subsection{2MASS}\label{2MASS}

The Two Micron All Sky Survey \citep[2MASS
\footnote{2MASS web site (http://www.ipac.caltech.edu/2mass/)},  ][]{Skrutskie2006-AJ}
is a project which observed 99.998\% of the whole sky 
at J (1.25 $\mu$m), H (1.65 $\mu$m), Ks (2.16 $\mu$m) bands, 
at Mt. Hopkins, AZ (the Northern Hemisphere) and at CTIO, Chile (the Southern Hemisphere)
between 1997 June and 2001 February. 
The instruments are both highly automated 1.3-m telescopes equipped with three-channel cameras, 
each channel consisting of a 256 $\times$ 256 array of HgCdTe detectors. 
The 2MASS obtained 4,121,439 FITS images (pixel size $\sim2''_{\cdot}0$) with 7.8 s of integration time.
The Point Source Catalogue (PSC) was produced using these images and catalogued 470,992,970 sources.
In the 2MASS web site, the images and the PSC are open to the public and are easily available.

\subsection{Overlapping Regions}

Overlapping regions are the edge parts of an image which overlap with neighboring images.
They are produced with the purpose of observing the whole sky without spatial gaps.
All of the 2MASS images have overlapping regions in their edges. 
The area of an overlapping region is larger than at least 10\% of an image area
and it increases at the higher declination.

We focused on these overlapping regions of the 2MASS images.
The existence of an overlapping region means the existence of a region 
where observed multiple times in different epochs.
Therefore, it is possible to detect variable objects by comparing the brightness of the objects 
in the overlapping regions.

\subsection{Photometric Accuracy \label{PA}}

The 2MASS PSC lists only single-epoch magnitude for an object 
even if it is in an overlapping region. 
In other words, we can know only a single-epoch magnitude from a 2MASS PSC source.
Therefore, in order to investigate the light variation of an object, 
we measured the relative magnitudes of objects in overlapping regions by ourselves.

Photometry was carried out with the APPHOT package in IRAF (Image Reduction and Analysis Facility
\footnote{IRAF is a software system for the reduction and analysis of astronomical data, 
which is distributed by the National Optical Astronomy Observatory.}) software. 
As mention in Sect. \ref{Detection-Criteria}, we subtract the objects 
which is contaminated by other objects (i.e., not single stars). 
Therefore, we used constant aperture size photometry 
to perform automated photometry easily. 
However, photometric accuracy may be worse 
when we perform photometry of an object in an overlapping region, 
since they exist at the edges of images.
Below, we discuss the validity of these measurements. 

\begin{figure*}[htbp]
  \begin{center}
    \begin{tabular}{ccc}
      \resizebox{50mm}{!}{\includegraphics[]{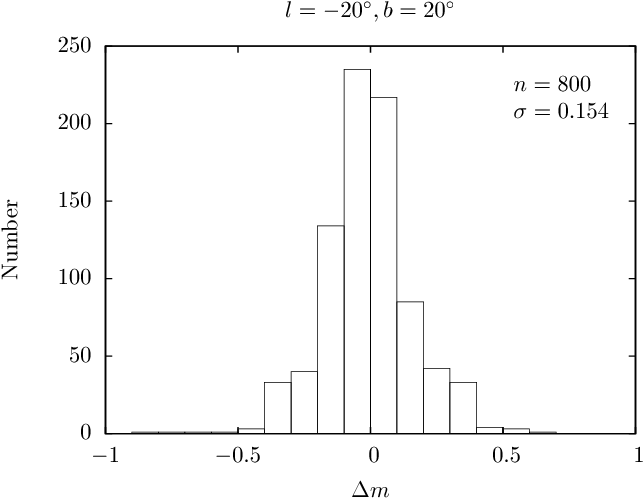}} &
      \resizebox{50mm}{!}{\includegraphics[]{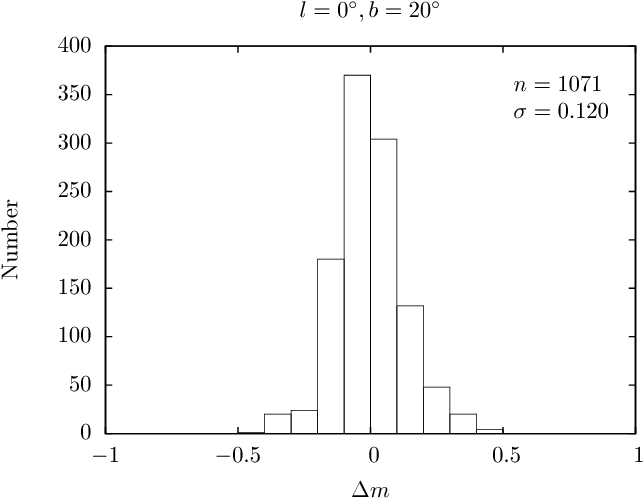}} &
      \resizebox{50mm}{!}{\includegraphics[]{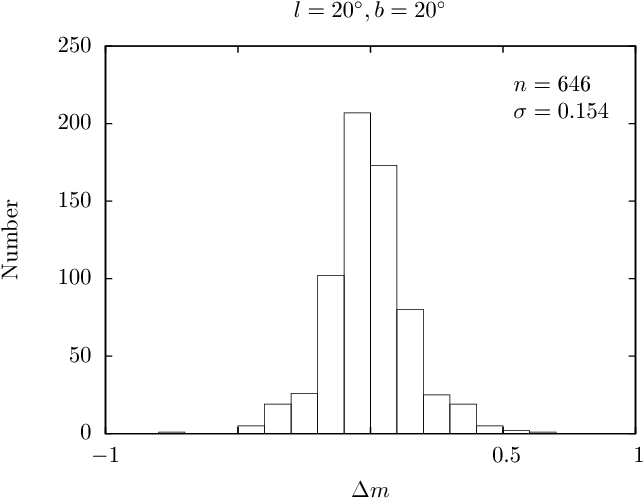}} \\
      \resizebox{50mm}{!}{\includegraphics[]{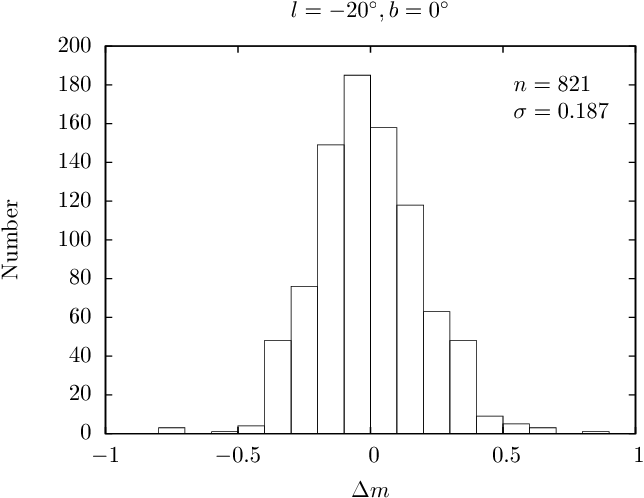}} &
      \resizebox{50mm}{!}{\includegraphics[]{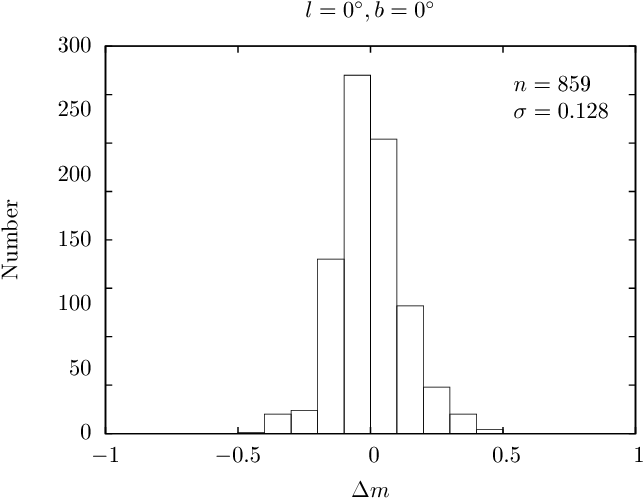}} &
      \resizebox{50mm}{!}{\includegraphics[]{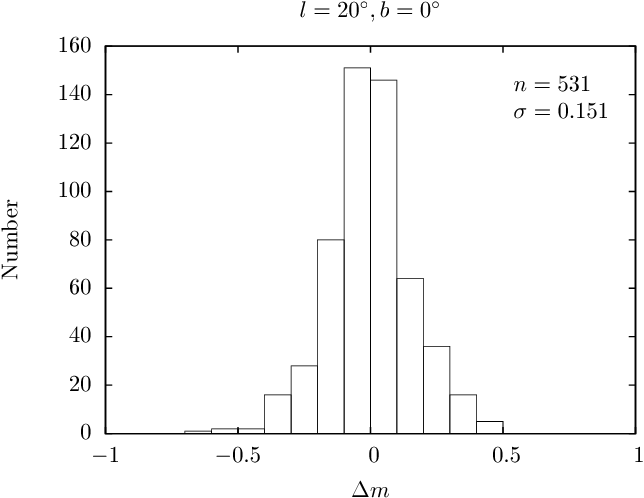}} \\
      \resizebox{50mm}{!}{\includegraphics[]{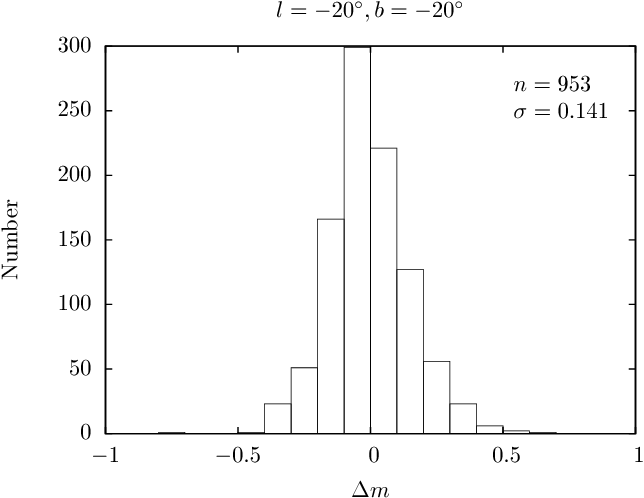}} &
      \resizebox{50mm}{!}{\includegraphics[]{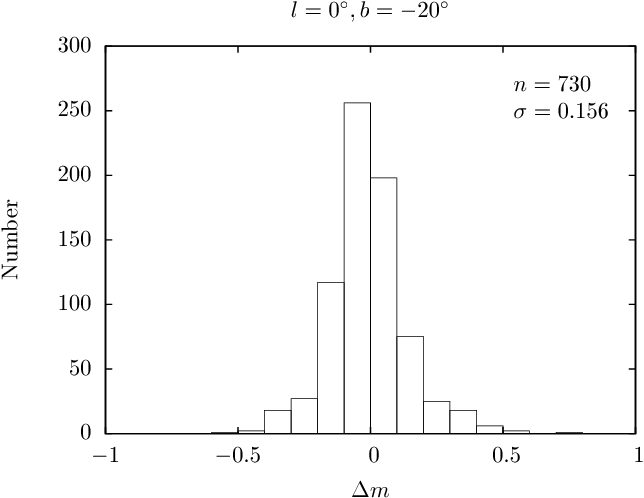}} &
      \resizebox{50mm}{!}{\includegraphics[]{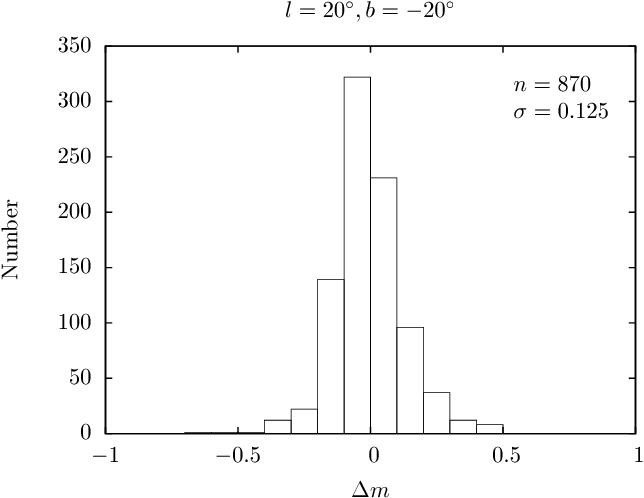}} \\
    \end{tabular}
    \caption{Histograms of the differences between our measurements and 2MASS magnitudes. 
			The number of samples and standard deviation 
            for the difference are denoted by $n$ and $\sigma$, respectively. 
			These areas are located at 
			$(l,b)=(-20^\circ, 20^\circ), (0^\circ, 20^\circ), (20^\circ, 20^\circ)$ (top panel), 
			       $(-20^\circ, 0^\circ), (0^\circ, 0^\circ), (20^\circ, 0^\circ)$ (middle panel),
				   $(-20^\circ, -20^\circ), (0^\circ, -20^\circ), (20^\circ, -20^\circ)$ (bottom panel).
}
    \label{photometric accuracy}
  \end{center}
\end{figure*}

The photometric accuracy was investigated on the basis of differences 
between our measurements and 2MASS PSC magnitudes (i.e., $\Delta m$ in Fig. \ref{photometric accuracy}). 
We selected 9 areas of which source densities are relatively higher and lower ($b=-20^\circ, 0^\circ, 20^\circ$).
The number of sample objects in each area is more than 500 objects.
In order to avoid the influence of inaccurately catalogued values, 
we used objects having photometric quality flags in the 2MASS PSC 
better than B (corresponding to SNR$>$7) in the J-band. 
This sensitivity limit corresponds to approximately 16.5 mag in the J-band 
\citep[see ][]{Skrutskie2006-AJ,Cutri2003-book}). 
It should be noted that we investigate the sources with $8 < J < 17$. 
The zero-point of our measurements was decided by 
making the average magnitude agree with the 2MASS average magnitude, 
and we calculated the standard deviation ($\sigma$) for the difference. 
Note that we show results only in the J-band 
because we used only J-band images to search for variable stars in this paper. 

Figure \ref{photometric accuracy} shows the histograms of the differences 
between our measurements and 2MASS magnitudes. 
Their distributions are within $\pm 0.5$ mag with $\sigma \sim 0.15$. 
We note that each $\sigma$ does not always represent the spread of the distribution 
because a part of their distributions is difficult to approximate to a gaussian distribution. 
The average $\sigma$ among the 9 areas is 0.146.
The differences of $\sigma$ among the 9 areas can be explained 
as increases in cases of including many extended sources or double-stars 
because we adopted constant aperture size.
Source density less affects our photometry 
because the differences of $\sigma$ among areas whose source densities are 
relatively high ($b=0^\circ$) and low ($b=20^\circ,-20^\circ$) are small. 
Few objects have $\Delta m \geq 3\sigma$ in all areas, 
so it is highly probable that such objects are real variable objects.
Accordingly, $\Delta m$ larger than 3$\sigma$ can be a reliable criterion to select variable objects. 
As describes in Sect. \ref{Detection-Criteria}, 
we extracted the sources having magnitude variance over $3 \sigma$ as variable star candidates. 

\begin{figure}[htbp]
\begin{center}
\includegraphics[clip]{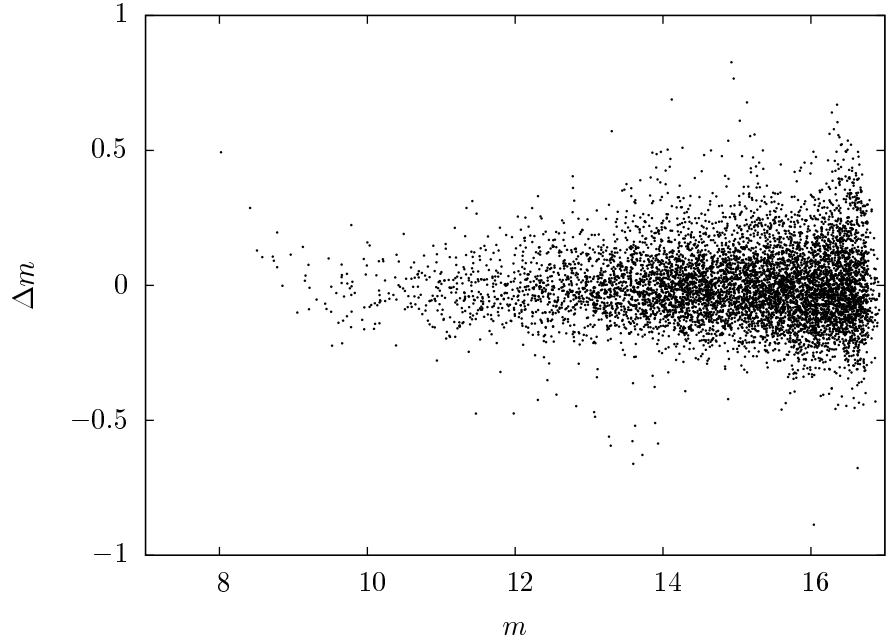}
\caption{The relation between 2MASS magnitude ($m$) and 
the differences between our measurements and 2MASS magnitudes ($\Delta m$). 
This comparison is performed at J-band. 
All of the samples are same as used in Fig. \ref{photometric accuracy}. 
  \label{mag-deltamag}}\end{center}
\end{figure}

Figure \ref{mag-deltamag} shows $m$-$\Delta m$ diagram. 
It presents that the both ends of each distribution in Fig. \ref{photometric accuracy} are 
the sources whose brightnesses are near the limiting magnitude. 
If the brightnesses of most stars in an overlapping region are near the limiting magnitude, 
it might be difficult to detect variable stars in the region because the $\sigma$ should be relatively large 
compared with the other overlapping regions. 

In this search, we calculated $\sigma$ between two images which have a common overlapping region, 
and picked up the objects with $\Delta m \geq 3 \sigma$. 
This procedure is performed on each overlapping region. 
It should be noted that the 10$\%$ of both the brightest and the faintest sources 
are not used for calculating $\sigma$ 
because photometric accuracy for these objects is considered to be worse. 
Most of the standard deviations are smaller than the value derived in the previous paragraph (i.e., 0.146), 
which are among approximately 0.1--0.15.

\section{Data Analysis \label{Data-Analysis}}

\subsection{Detection Procedure}

In order to detect variable objects, we analyzed 2MASS public images according to the following steps.

\begin{enumerate}
\item Identify the overlapping region between neighboring images. 
\item Perform relative photometry for the objects in the overlapping region. 
\item Compare the relative magnitudes. 
\item Confirm the variabilities in public images by eye. 
\end{enumerate}

Step(1): At first, the overlapping region should be identified. 
The sizes of overlapping regions are different in each image, so they have to be identified individually.
We identified the overlapping region by moving the other image little by little 
until both positions of most objects agree with each other.

Step(2): 
The 2MASS public images were subjected to standard processing, i.e. 
they were corrected for instrumental signatures 
by subtracting a dark frame, dividing by a responsibility image (flat-field), 
and subtracting a sky-illumination correction image \citep{Skrutskie2006-AJ}.
Therefore, they can be used for photometry as they are.
We measured magnitudes by aperture photometry with the APPHOT package in the IRAF.

Step(3): 
We investigated the variability of an object by comparing the relative magnitudes derived in step(2), and 
picked up the objects satisfying detection criterion(1) in Sect. \ref{Detection-Criteria}.

Step(4): 
In the case of double-stars seen in Fig. \ref{photo}(a), 
photometric accuracy might be considerably worse because of performing aperture photometry.
In order to remove such kinds of objects and detect variable objects with high reliability, 
we confirmed the variabilities by examining 2MASS public images.
Finally, we extracted the objects satisfying detection criterion(2) in Sect. \ref{Detection-Criteria}.

It is not realistic to perform this procedure by hand due to the vast amount of images, 
so we produced a semi-automated system for performing these steps.
In this procedure, image reduction was carried out with the IRAF software package and 
the other parts were carried out with Fortran90.

\subsection{Detection Criteria \label{Detection-Criteria}}

The detected objects in this paper satisfy the following criteria.

\begin{figure*}[htbp]
\begin{center}
\begin{tabular}{cc}
(a) & (b) \\
\begin{minipage}[htbp]{0.5\textwidth}
\begin{center}
	\includegraphics[width=2cm,clip]{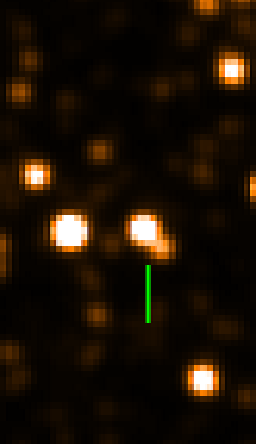}
	\hspace{10mm}
	\includegraphics[width=2cm,clip]{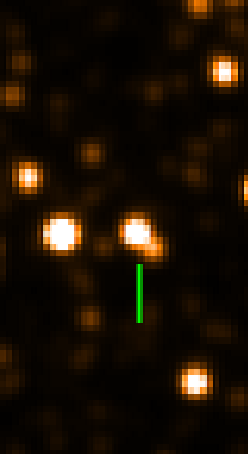}
\end{center}
\end{minipage}
 & 
\begin{minipage}[htbp]{0.5\textwidth}
\begin{center}
	\includegraphics[width=2cm,clip]{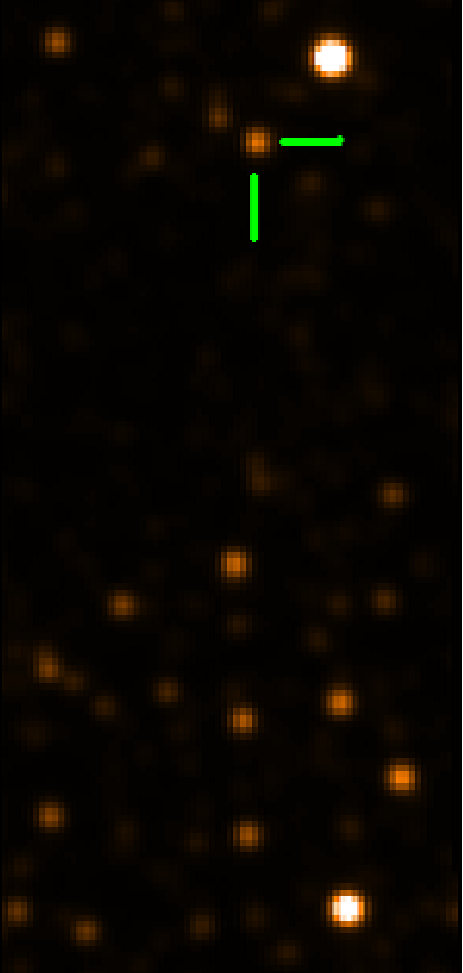}
	\hspace{10mm}
	\includegraphics[width=2cm,clip]{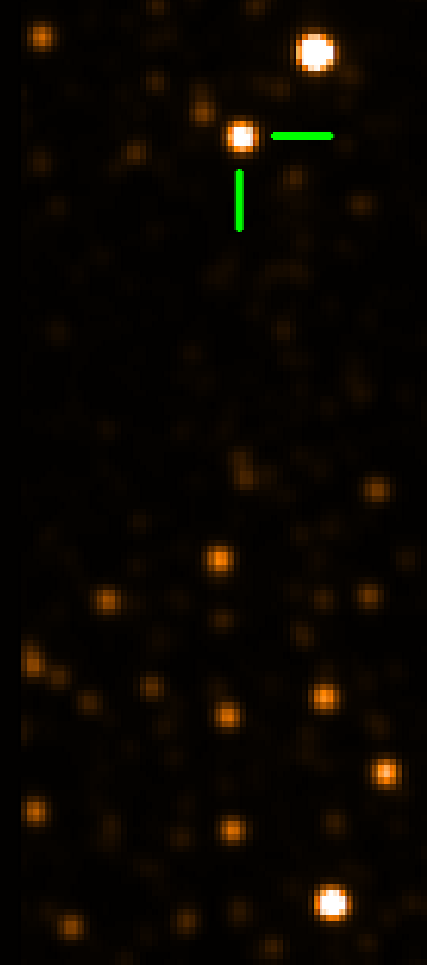}
	\label{detected-star}
\end{center}
\end{minipage}
\end{tabular}
\caption{(a)An example of double-stars. In this case, the accuracy of aperture photometry might be sub-optimal.
(b)Actual detected objects as variable objects. 
While the other sources looks like same brightness in the images, 
the brightness of checked source is clearly different.
The difference of magnitudes is approximately 1.3 mag at the J-band.
\label{photo}}
\end{center}
\end{figure*}

\begin{enumerate}
\item Variability over 3$\sigma$

As discussed in Sect. \ref{PA}, 
variability larger than $3\sigma$ is a reliable criterion to judge whether a source is variable object or not 
despite the use of sources in the edge parts of images 
where the photometric accuracy might be sub-optimal.
Hence, we picked up the objects as variable star candidates which have variability larger than $3\sigma$.

\item Objects showing apparent difference in images

Most of the objects picked up on the basis of criterion(1) 
are actually a confusion of double-stars for single stars (see Fig. \ref{photo}(a)).
In these cases, photometric accuracy might be worse or the counterpart might not be able to be extracted accurately.
In order to extract more reliable candidates, we confirmed the variability by examining images.
After we selected clearly single stars, finally, 
we extracted only sources whose brightness is obviously different compared with the other sources (see Fig. \ref{photo}(b) ).
\end{enumerate}

In this paper, we examined only overlapping regions having interval of epochs 
larger than one day because of following reasons. 
Most of the detected variable stars are expected to be red giant variables 
because we investigate in the direction of the Galactic centre, 
and the criterion(1) will extract many sources with $\Delta J > \sim 0.4$. 
The red giant variables are considered to hardly have such large light variance in one day. 
Therefore, few red giant variables are expected to be detected in the overlapping regions with one-day interval of epochs. 
Additionally, there is a large number of images with one-day interval of epochs, 
so it is difficult to examine all of these images by eye. 
Accordingly, we examined only overlapping regions having interval of epochs larger than one day.

\subsection{Extraction of variable stars}

\begin{figure}[thbp]
\begin{center}	
\plotone{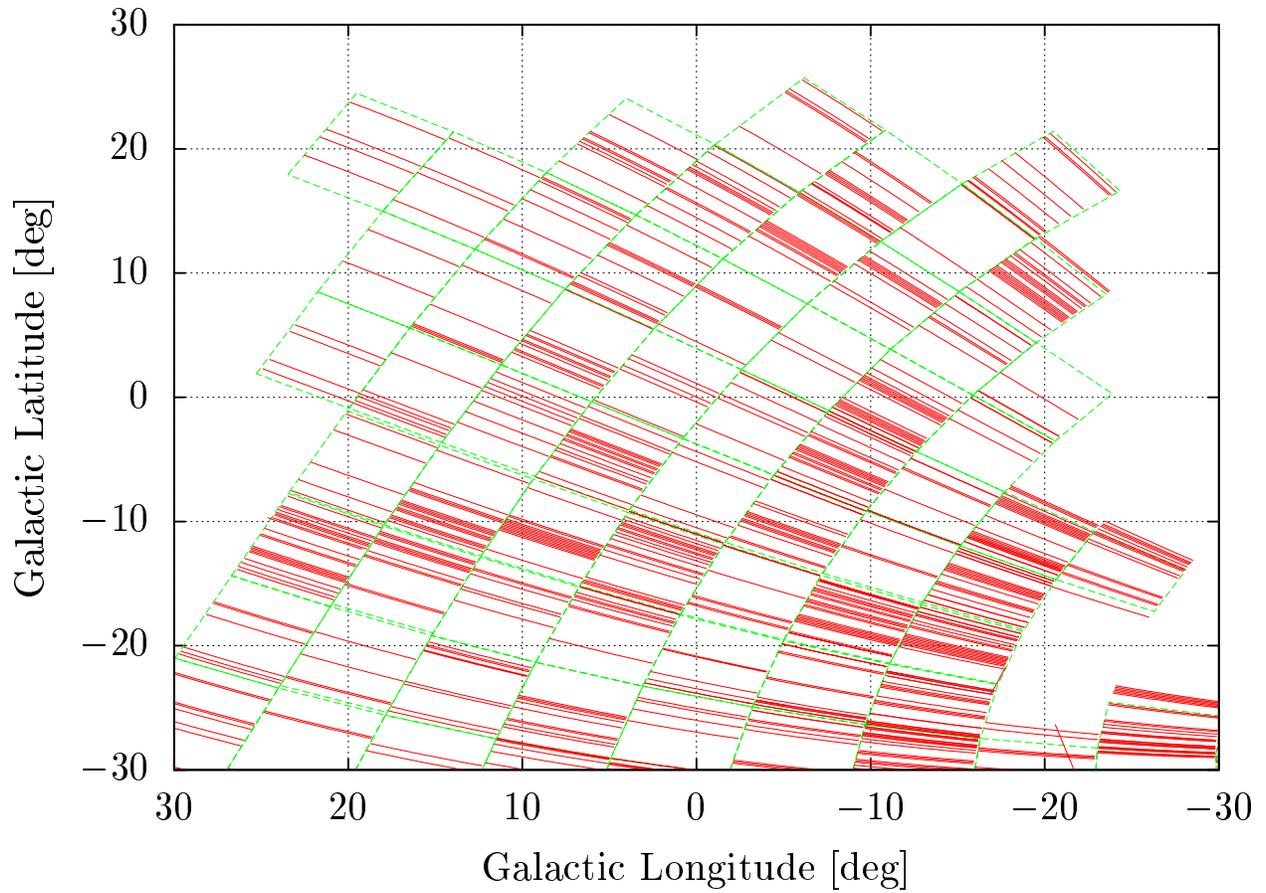}
\caption{Investigated region ($-30^\circ \lesssim l \lesssim 20^\circ, |b|\lesssim 20^\circ$) 
				are shown by the red solid lines themselves. 
 			We have only investigated the regions whose interval of epochs are larger than one day.
 			The green broken frames indicate $7^\circ \times 6^\circ$ regions.
 \label{Investigated-Region}}
\end{center}
\end{figure}

Figure \ref{Investigated-Region} shows the investigated region, which covers the entire bulge 
($-30^\circ \lesssim l \lesssim 20^\circ, |b| \lesssim 20^\circ$). 
Variable stars generally have smaller amplitudes in H or K$_\textnormal{\tiny S}$ bands than those in J-band. 
Therefore, many variable stars are expected to be detected at J-band compared with at H or K$_\textnormal{\tiny S}$ bands. 
In addition, even if we can detect variable stars only at H and K$_\textnormal{\tiny S}$ bands 
using H or K$_\textnormal{\tiny S}$ images, 
we can not derive the information of the detected variable stars at J-band. 
Hence, we used only J-band images to search for variable stars in this paper. 
The number of the investigated images is $\sim$15 000 
(corresponding to approximately 1$\%$ of the whole sky). 
It should be noted that each area of the overlapping regions in this search 
occupies approximately $20 \pm 4\%$ of an image. 
The total area of the investigated overlapping regions is roughly $100$ deg$^2$. 

As a result of the investigation, we have detected 136 variable objects. 
The catalogue of the detected variable stars are shown in Table \ref{catalogue}. 
Table \ref{detailed-bulge} shows the number of the detected objects 
in each divided area towards the Galactic centre 
together with the number of investigated images. 

Investigating whether the detected objects are already known using VizieR Service 
\footnote{VizieR Service is a database of astronomical catalogues, 
which is provided by the Centre de Donn\'ees astronomiques de Strasbourg (CDS), France.} \citep{Ochsenbein2000}, 
we found only six objects are listed in the variable star catalogues shown in Table \ref{VS-catalogues}.

\subsubsection{Optical counterpart}
We examined Digitized Sky Survey (DSS) images by eye to check the optical counterparts of our sample. 
As a result, 84 objects were confirmed as very faint sources. 
However, almost all of them have no magnitudes in the USNO-B1.0 catalogue \citep{Monet2003} 
because they are too faint to measure magnitudes.
The existence of optical counterparts are also listed in our catalogue.

\subsection{Interstellar Extinction}
Near the Galactic centre, the interstellar extinction is considerably large 
due to a large amount of dust. 
Accordingly, even in the near-infrared wavelength, 
the extinction can not be ignored. 
\citet{Dutra2003-MNRAS} estimated the interstellar extinction 
in $|l|\leq6^\circ, |b|\leq6^\circ$ region. 
However, we detect variable stars in $|l|\leq20^\circ, |b|\leq6^\circ$ region. 
Hence, we estimated the extinction of the detected variable stars 
using a method similar to \citet{Dutra2003-MNRAS}, 
which use a near-infrared Colour-Magnitude Diagram (CMD). 
It should be noted that we do not exclude extinction by circumstellar dust. 
First, we collected stars distributed within $10'$ of the position of each detected variable star, 
and measured the shift of the upper giant branch from the intrinsic position 
along the reddening vector on the $(J-K_\textnormal{\tiny S})$-$K_\textnormal{\tiny S}$ CMD. 
Both the intrinsic position of RGB and the reddening vector, as follows, 
are taken from \citet{Dutra2002-AA}.
\begin{eqnarray}
(K_\textnormal{\tiny S})_0 = -7.81 (J-K_\textnormal{\tiny S})_0 + 17.83 \\
A_\textnormal{\tiny K} / E(J-K) = 0.670
\end{eqnarray}
As mentioned in \citet{Dutra2003-MNRAS}, the extinction for the lower latitude is underestimated 
by contamination of many blue stars with $K_\textnormal{\tiny S} \lesssim 11$. 
Thus, at the lower latitude ($b \leq 5^\circ$) we removed the stars 
with $K_\textnormal{\tiny S}<11.5$ as performed in \citet{Dutra2003-MNRAS}.

\begin{figure}[htbp]
		\plotone{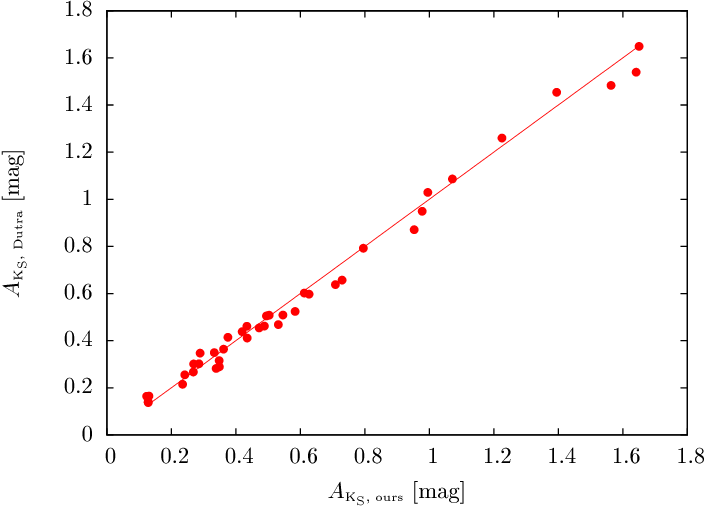}
		\caption{Comparison of $A_\textnormal{\tiny K$_\textnormal{\tiny S}$}$ 
for 40 fields between \citet{Dutra2003-MNRAS} and our estimation. 		\label{comparison-of-Ak}}
\end{figure}

The extinction values are given in our catalogue. 
A part of these values are compared with the result of \citet{Dutra2003-MNRAS} 
in order to examine the accuracy of our estimation. 
As shown in Fig. \ref{comparison-of-Ak}, our estimations agree well with that of \citet{Dutra2003-MNRAS}. 
The standard deviation for the difference of $A_\textnormal{\tiny K$_\textnormal{\tiny S}$}$ is $ 0.04$ mag. 

Although the metallicity variations in the bulge might make our estimation inaccurate, 
we ignored them because it is considered to be small \citep[e.g., ][]{Frogel1999-AJ,Ramirez2000-AJ}.

\section{Detection probability based on twice-epoch magnitudes \label{Probability}}

As mentioned in Sect. \ref{2MASS}, 
the 2MASS images were obtained in four years survey. 
An interval of epochs in an overlapping region is not always same as that in the other overlapping region. 
In addition, we can compare only twice-epoch magnitudes to search for variable stars. 
Accordingly, the difference among intervals of epochs might cause our selection to be biased one, 
although our selection is expected to be unbiased because we investigated a large number of sources. 
Therefore, we investigated detection probability based on twice-epoch magnitudes 
by performing Monte Carlo simulation under some simple assumptions.

\subsection{Modelling}
Although there are a large number of SRVs in the Galactic bulge, 
the detected variable stars should be Mira variables or OH/IR stars as show in the later section. 
Their light curves can be assumed by a sinusoid. 
Therefore, we consider the region that contains $N=100$ variable stars 
whose light curves obey a symmetric sinusoidal model, 
\begin{eqnarray}
m=\frac{A}{2} \sin \left( 2 \pi \frac{\Delta d}{P}+\theta_0 \right)
\end{eqnarray}
where $m$ is the magnitude, $A$ is the amplitude, 
$\Delta d$ is the interval of epochs in an overlapping region, 
$P$ and $\theta_0$ are the period and the initial phase of a variable star. 
We set $A$ $(0 \leq A \leq 2)$ and $\theta_0$ $(0 \leq \theta_0 \leq 2 \pi)$ as random parameters, 
and calculated the detection probability as a function of $\Delta d$. 
We consider that a period distribution for variable stars 
is a probability distribution of period, 
and a period for a variable star is randomly given on the basis of the probability distribution.

\subsection{Detection probability}\label{Detection-probability}
We considered two cases for the distribution of period: 
(a) uniform ($P \leq 1000$) and (b) gaussian distribution.
We assume the gaussian distribution having the average period of 350 days and the sigma of 100 days, 
which approximately reproduce the period distribution in \citet{Groenewegen2005-AA}. 
We set the detection criterion to be 0.45, which is approximately 
average $3\sigma$ level in our search (see Sect. \ref{PA}).

\begin{figure*}[tbp]
\begin{tabular}{cc}
(a) & (b) \\
	\begin{minipage}[htbp]{0.5\textwidth}
	\begin{center}
		\resizebox{90mm}{!}{\includegraphics[clip]{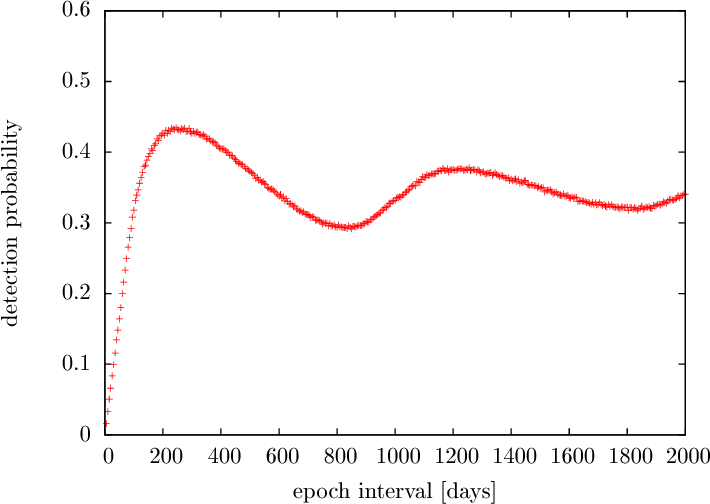}}
	\end{center}
	\end{minipage}
 &
	\begin{minipage}[htbp]{0.5\textwidth}
	\begin{center}
		\resizebox{90mm}{!}{\includegraphics[clip]{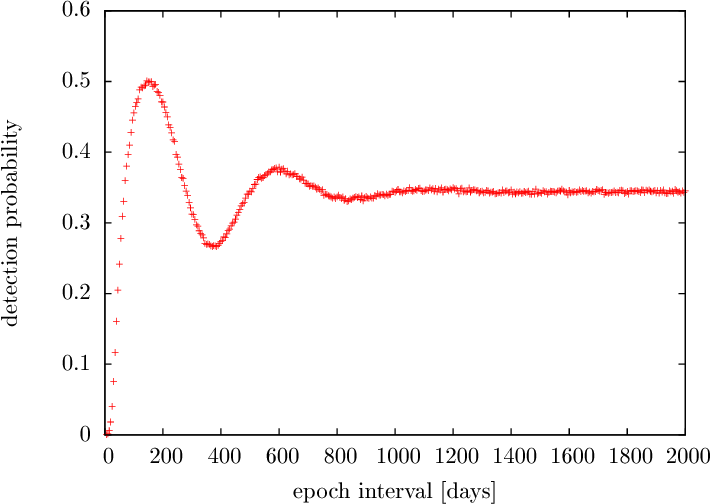}}
	\end{center}
	\end{minipage}
\end{tabular}
		\caption{The result of Monte Carlo simulations: (a) uniform and (b) gaussian distribution.  
		\label{dec-prob}}
\end{figure*}

The numerical results are shown in Fig. \ref{dec-prob}. 
Each plot is average value of 1000 times simulations. 
Whereas two curves are very similar in shape, 
the difference between them is the depth of both peak and trough. 
The range of detection probability in Fig. \ref{dec-prob}(a) is 
approximately 0.34 (average value) $\pm 0.1$, 
whereas that in Fig. \ref{dec-prob} (b) is nearly same as (a)
though it is as large as $\sim 0.5$ between 100 and 220 days. 
Hence, the detection probabilities can be considered to be 
less sensitive to interval of epochs within 0.1 detection probability. 

\citet{Schultheis2000-AA} presented two catalogues of variable star candidates 
in an area of $\sim 4$ deg$^2$ of the inner galactic bulge. 
They compared their catalogues with the catalogue of \citet{Glass2001-MNRAS}, and 
claim that $\sim 40$\% of variable stars can be recovered using only twice-epoch measurements.
Assuming the rms error to be 0.16 mag and detection criterion to be $2\sigma$ level 
as described in \citet{Schultheis2000-AA}, 
we can derive the average detection probability of $\sim 0.44 \pm 0.05$ (both period distributions). 
This is consistent with the claim of \citet{Schultheis2000-AA}. 

This simulation support that we can extract variable stars at partly constant probability 
independent on interval of epochs even if we compare only twice-epoch magnitudes for search. 
However, we note that the number of detected variable stars in this paper is considerably small 
due to our strict detection criteria (especially the criterion(2) ).


\section{Discussion}\label{Discussion}

\subsection{Detection Rate}

Here, we estimate an actual detection rate by comparing the result of \citet{Schultheis2000-AA}. 
\citet{Schultheis2000-AA} investigated $\sim 4$ deg$^2$ area in $-4^\circ \lesssim l \lesssim 1^\circ$, 
$-1^\circ \lesssim b \lesssim 1^\circ_\cdot 5$, and 
have detected $\sim 720$ variable stars with $2 \sigma$ level at J-band. 
When their second detection criterion (i.e., $\Delta J>2\sigma$) 
is substituted for our criterion (i.e., $\Delta J> 3\sigma$), 
the number of their variable stars becomes $\sim 500$ objects. 
Then the number density becomes $\sim 130$ counts deg$^{-2}$. 
Because their samples would include some spurious variable stars, 
the actual number density should be smaller than this value. 
On the other hand, we have detected 38 variable stars in an area of 2.3253 deg$^2$ within $|l|\leq 5^\circ$, 
$|b| \leq 5^\circ$ (see Table \ref{detailed-bulge}), 
then the number density becomes 16.3 counts deg$^{-2}$. 
The number density of \citet{Schultheis2000-AA} is roughly eight times larger than that of ours. 
However, most of variable star candidates detected by our criterion 1 is contaminated by other objects and 
we exclude such objects to extract reliable variable stars as mentioned above. 
Therefore, we can probably recover $\sim 10\%$ of variable stars under our strict criteria. 


\subsection{Spatial Distribution}\label{Disc-SD}

Figure \ref{SDwithGCVS} shows the spatial distribution of the detected variable stars. 
The red filled circles indicate the variable stars with an optical counterpart. 
Variable stars in the General Catalogue of Variable Stars (GCVS) are also plotted on the diagram. 

The 118 variables are distributed in the $|b| \leq 5^\circ$ region (i.e., the Galactic bulge). 
This can be explained because the Galactic bulge contains considerably large number of variable stars 
compared with the outside of the bulge. 

Most of the detected variable stars are distributed in the region where 
there is relatively small number of GCVS variables. 
The GCVS variables have been discovered in optical wavelength. 
Thus, the small number of GCVS variables are probably caused by observing obscured region. 
Almost all of the detected variable stars have not been listed in GCVS catalogue 
though nearly half of them have an optical counterparts. 
This is because most of them are very faint sources in the optical ($R> \sim18$). 
It can be interpreted that near-infrared search enables us to detect variable stars 
in the inner bulge where can not be observed in the optical because of a large amount of dust.

\subsection{Photometric Properties}

\subsubsection{Colour-Magnitude Diagram}\label{CMD}
The $(J-K_\textnormal{\tiny S})_0$-$K_{\textnormal{\tiny S}, 0}$
CMD is shown in Fig. \ref{CMD1}. 
The distribution of our variable stars is similar to that of variable stars in the Galactic bulge 
\citep[e.g., ][]{Schultheis2000-AA,Groenewegen2005-AA}.
Most of the variable stars are located above the tip of the bulge RGB 
\citep[$\sim 8.0$--$8.2$, ][]{Tiede1995-AJ,Frogel1999-AJ}. 
These stars correspond to Asymptotic Giant Branch (AGB) stars. 
Still there are a small number of variable stars below the RGB tip, 
although three variable stars with $K_0>8.0$ (shown by the triangles in Fig. \ref{CMD1}) 
are located in the outside of the bulge ($|b|\gtrsim10^\circ$). 
The other variable stars below the RGB tip might be less luminous AGB or highly obscured AGB. 
\subsubsection{Light Variance}\label{Light-Variance}
The long-period variable stars such as SRVs, Mira variables and OH/IR stars are 
believed to be on the AGB phase. 
The detected variable stars are expected to be such AGB variables 
because most of the variable stars are AGB stars as shown in Sect. \ref{CMD}. 
Among these AGB variables, the SRVs have relatively small amplitudes 
though there are a few number of large-amplitude SRVs. 
Figure \ref{mag-amp} shows $J$-$\Delta J$ and $J_0$-$\Delta J$ diagrams for the detected variable stars. 
We note that $\Delta J$ does not represent amplitude, but represents minimum limit of the amplitude. 
Most of our samples have large light variance of $\Delta J > 0.5$. 
Therefore, it is highly probable that almost all of the variable stars are large-amplitude variable stars 
such as Miras and OH/IR stars. 

Figure \ref{mag-amp} also presents that 
the number of the detected variable stars with an optical counterpart extremely decreases in $J>11$. 
Additionally, most of the detected variable stars with $J>11$ have 
dereddened J magnitudes smaller than 11 mag (i.e., $J_0<11$). 
Hence, these are not intrinsically faint but are faint caused by interstellar absorption.

\subsubsection{Colour-Colour Diagram}
A near-infrared CCD is a powerful tool to investigate photometric properties of objects, 
and is well studied the loci of various objects. 
Figure \ref{CCD1} shows the $(H-K_\textnormal{\tiny S})$-$(J-H)$ and 
$(H-K_\textnormal{\tiny S})_0$-$(J-H)_0$ diagrams. 
The loci of Me-Mira (taken from \citealt{Kerschbaum2001-AA} compiling \citealt{Feast1982-MNRAS}), 
red SRVs and blue SRVs \citep{Kerschbaum1996-AAS} are also shown in Fig. \ref{CCD1}. 
The reddening vector was taken from \citet{Rieke1985-ApJ} and 
shown as an arrow corresponding to 10 mag extinction at V-band. 
The colours are transformed from ESO and SAAO to 2MASS photometric system on the basis of \citet{Carpenter2001-AJ}. 
We note that similar distribution for Miras are also derived by some other studies 
\citep[e.g., ][]{Whitelock1994-MNRAS,Glass1995-MNRAS,Groenewegen2005-AA,Matsunaga2005-MNRAS}.

Most of the $(H-K_\textnormal{\tiny S}) \gtrsim 1.3$ variables without an optical counterpart 
have the dereddened colour of $(H-K_\textnormal{\tiny S})_0 \lesssim 1.3$. 
These stars are considered to be reddened by 
not circumstellar dust but a large amount of interstellar dust in the inner bulge. 

Our samples are distributed around the locus of the Me-Mira or reddened region from the locus, 
whereas few samples are distributed in the region reddened from the loci of red and blue SRVs. 
In other words, few samples are SRVs. 
This is consistent with the property of light variance as mentioned in Sect. \ref{Light-Variance}. 
The Mira-like SRV is also a possible object, 
but the detected variable stars should not be the Mira-like SRVs 
because their locus on a near-infrared CCD is similar to that of red SRVs \citep{Kerschbaum1994-AAS}.  

Although both Mira variables and OH/IR stars are long-period and large-amplitude variable stars, 
the majority of OH/IR stars near the Galactic centre 
have colour larger than $(H-K)_0 \sim 1.0$ \citep{Wood1998-AA}. 
These stars are distributed over a wide range of the near-infrared colour 
extending toward redder colours of the region where optically visible Mira variables normally found. 
This is mainly due to undergoing heavy mass loss. 
Though OH/IR stars would be faint due to heavy circumstellar dust, 
they can be detected in both optical and near-infrared wavelengths \citep{Wood1998-AA,Jimenez2005-AA}. 
Our samples show that the colours of $\sim 80$\% variable stars to be $(H-K_\textnormal{\tiny S})_0 <1 $. 
Hence, it is highly probable that most of the detected variable stars are Mira variables 
though there are a small fraction of OH/IR stars. 
It should be noted that three objects in our samples have been detected as OH or SiO maser sources. 
Table \ref{maser-catalogues} shows the catalogues listing up the three objects.

\subsection{Number density distribution}

\begin{figure}[tbp]
	\begin{center}
		\resizebox{\hsize}{!}{\includegraphics[clip]{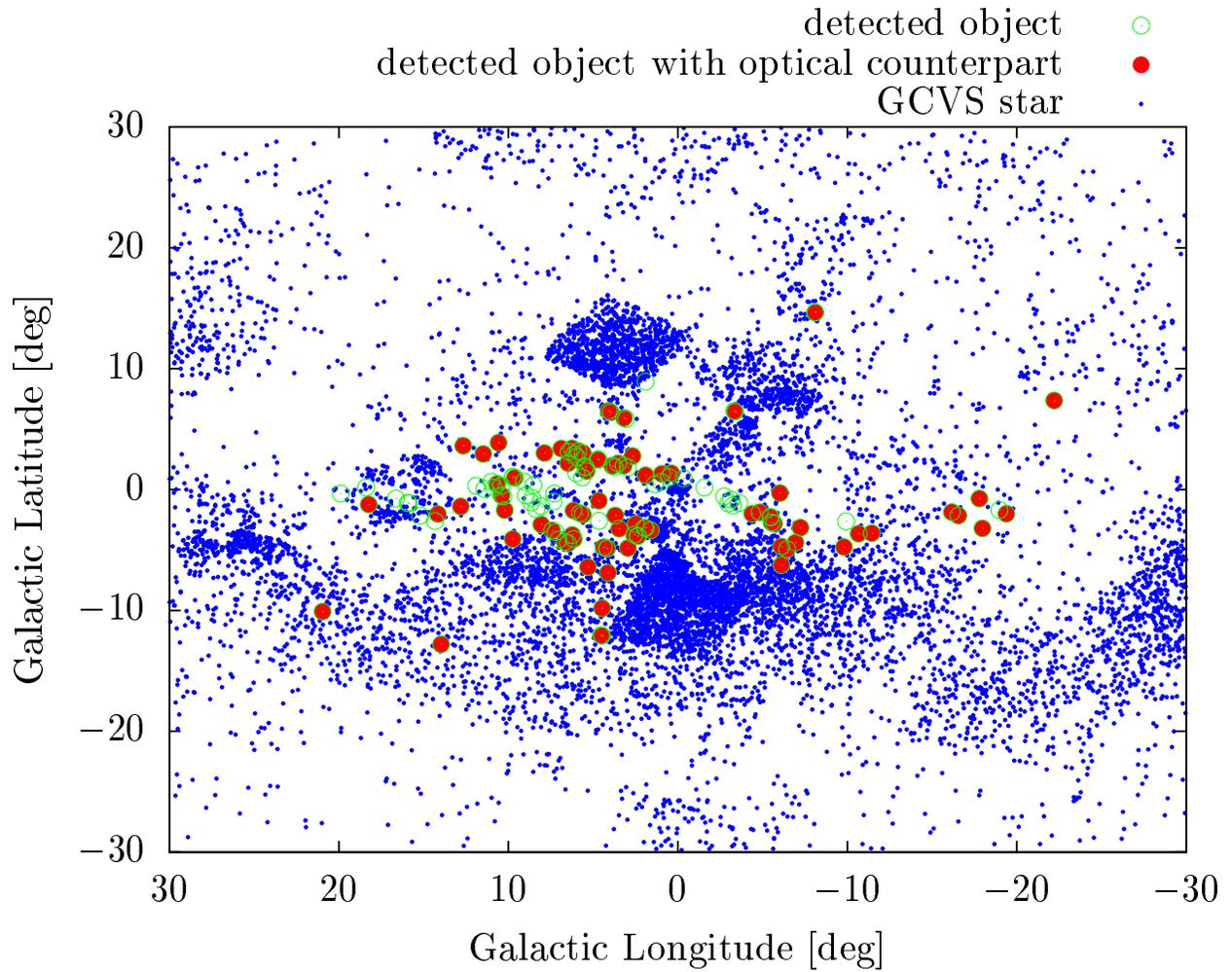}}
		\caption{Spatial distribution for the detected variable stars in the Galactic coordinate. 
			      The samples of already-known variable stars are also plotted on the diagram.
				  They were extracted from 
					Combined General Catalogue of Variables Stars \citep{Samus2004-catalogue}.
		\label{SDwithGCVS}}
	\end{center}
\end{figure}

\begin{figure}[tbp]
	\begin{center}
		\resizebox{\hsize}{!}{\includegraphics[clip]{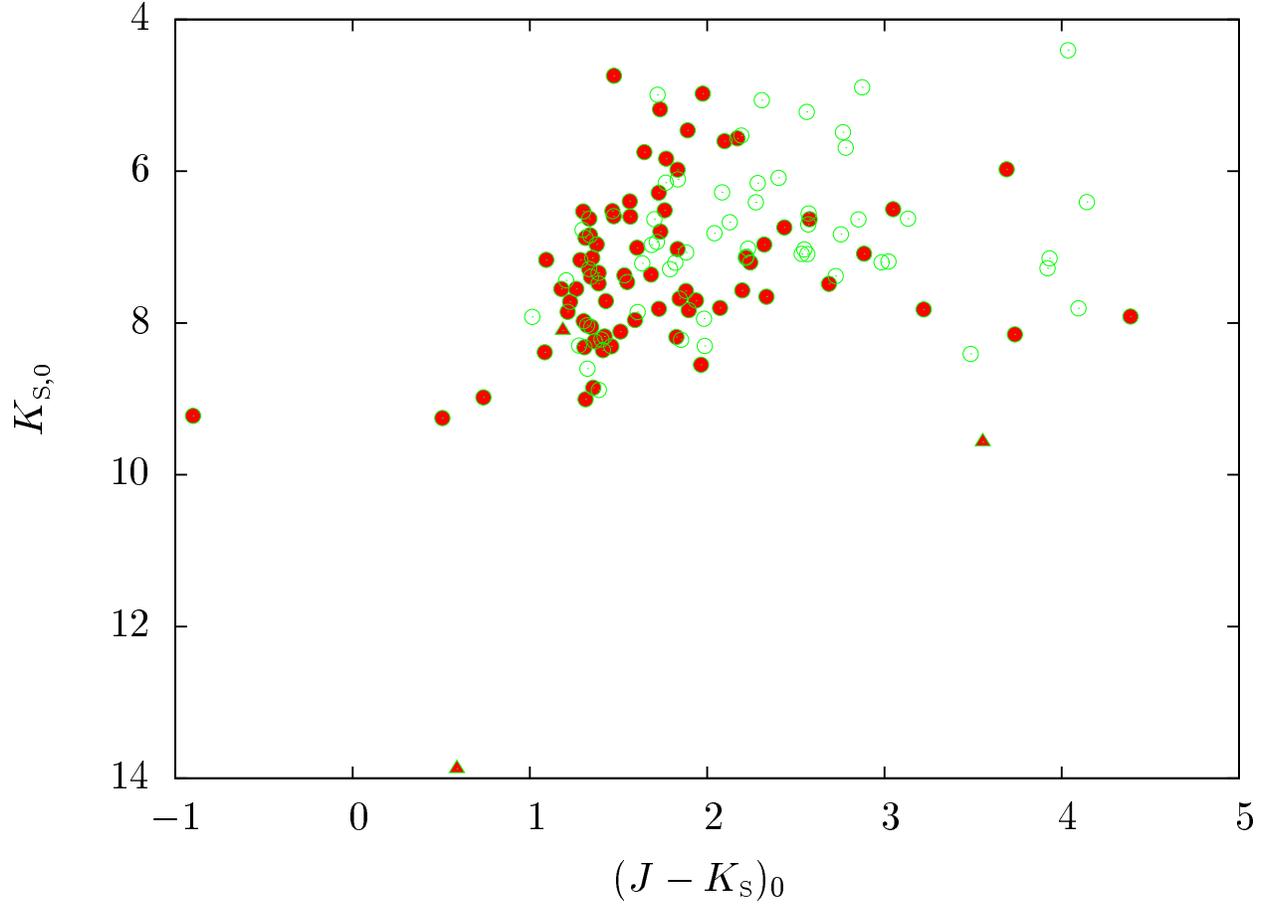}}
		\caption{Colour-magnitude diagram for the dereddened colour. 
				The symbols are same as in Fig. \ref{SDwithGCVS}. 
				The triangles show variable stars with $K_0>8.0$ in the outside of the bulge ($|b|\gtrsim10^\circ$, 
see Sect. \ref{CMD}).
		\label{CMD1}}
	\end{center}
\end{figure}

\begin{figure}[tbp]
	\begin{center}
		\resizebox{150mm}{!}{\includegraphics[clip]{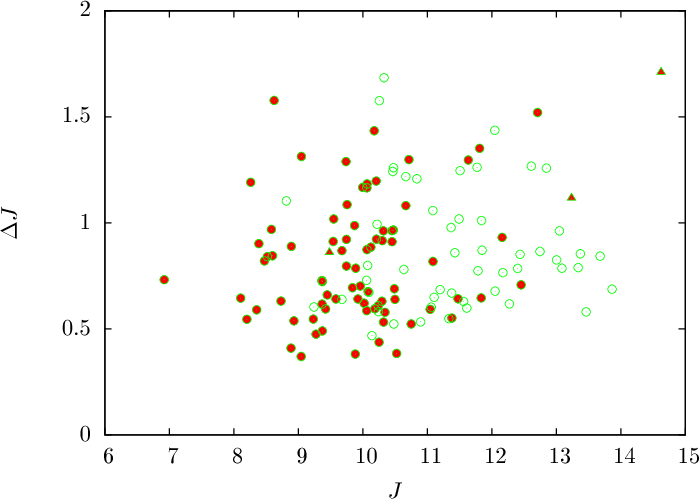}}
		\resizebox{150mm}{!}{\includegraphics[clip]{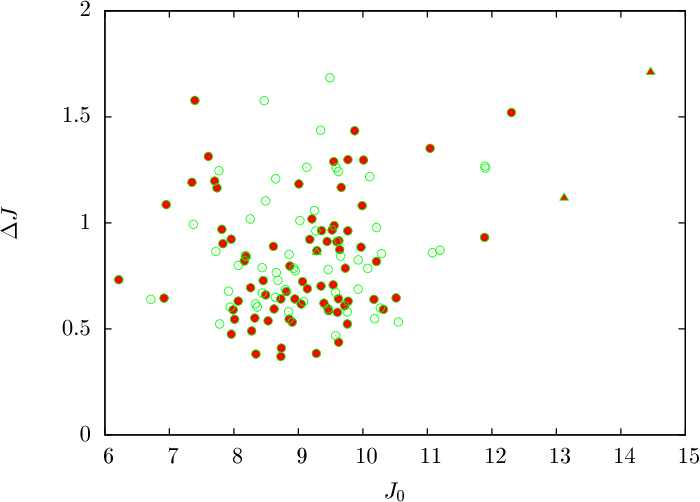}}
		\caption{$J$-$\Delta J$ (top panel) and $J_0$-$\Delta J$ (bottom panel) diagrams for the detected variables. 
				The $\Delta J$ is the light variance between two epoch measurements. 
				The symbols are same as in Fig. \ref{CMD1}. 
		\label{mag-amp}}
	\end{center}
\end{figure}

\begin{figure*}[htbp]
	\begin{center}
		\resizebox{100mm}{!}{\includegraphics[clip]{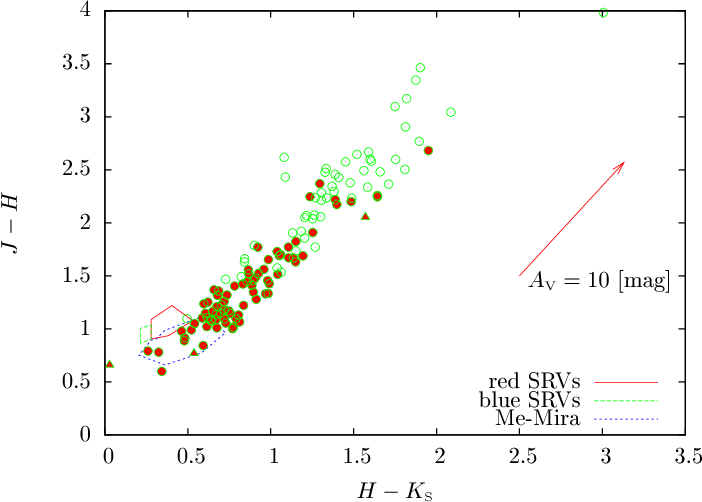}}
		\resizebox{100mm}{!}{\includegraphics[clip]{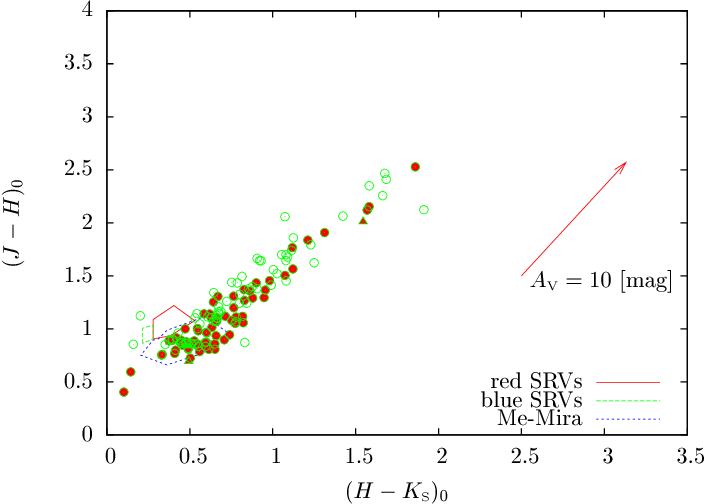}}
		\caption{Colour-colour diagrams for the detected variable stars. 
The symbols are same as in Fig. \ref{CMD1}. 
The red and green line boxes indicate the regions of red and blue SRVs taken from \citet{Kerschbaum1996-AAS}. 
Mira-like SRVs are also distributed in the region same as that of red SRVs \citep{Kerschbaum1994-AAS}.
The blue broken box indicates the distribution of Mira variables taken from \citet{Kerschbaum1996-AAS} 
that compiles the data from \citet{Feast1982-MNRAS}. 
		\label{CCD1}}
	\end{center}
\end{figure*}

In Fig. \ref{SDwithGCVS}, we can notice the number of the detected variable stars 
in $l > 0^\circ$ is larger than that in $l< 0^\circ$. 
However, we can not compare the number of variable stars between $l>0^\circ$ and $l<0^\circ$ in Fig. \ref{SDwithGCVS} 
because the number of investigated images is different in each region (see Fig. \ref{Investigated-Region}). 
We then divided $|l| \leq 20^\circ$, $|b| \leq 5^\circ$ region into 
$\delta l \times \delta b = 1^\circ \times  10^\circ$ areas, 
and calculated the number density of detected variable stars every $1^\circ$ Galactic longitude 
(i.e., in a $\delta l \times \delta b = 1^\circ \times  10^\circ$ bin, 
the area of whole region is divided by the area of investigated region, 
which is additionally multiplied by the number of detected variable stars in each bin).

\begin{figure}[thbp]
	\begin{center}
		\resizebox{\hsize}{!}{\includegraphics[width=6cm,clip]{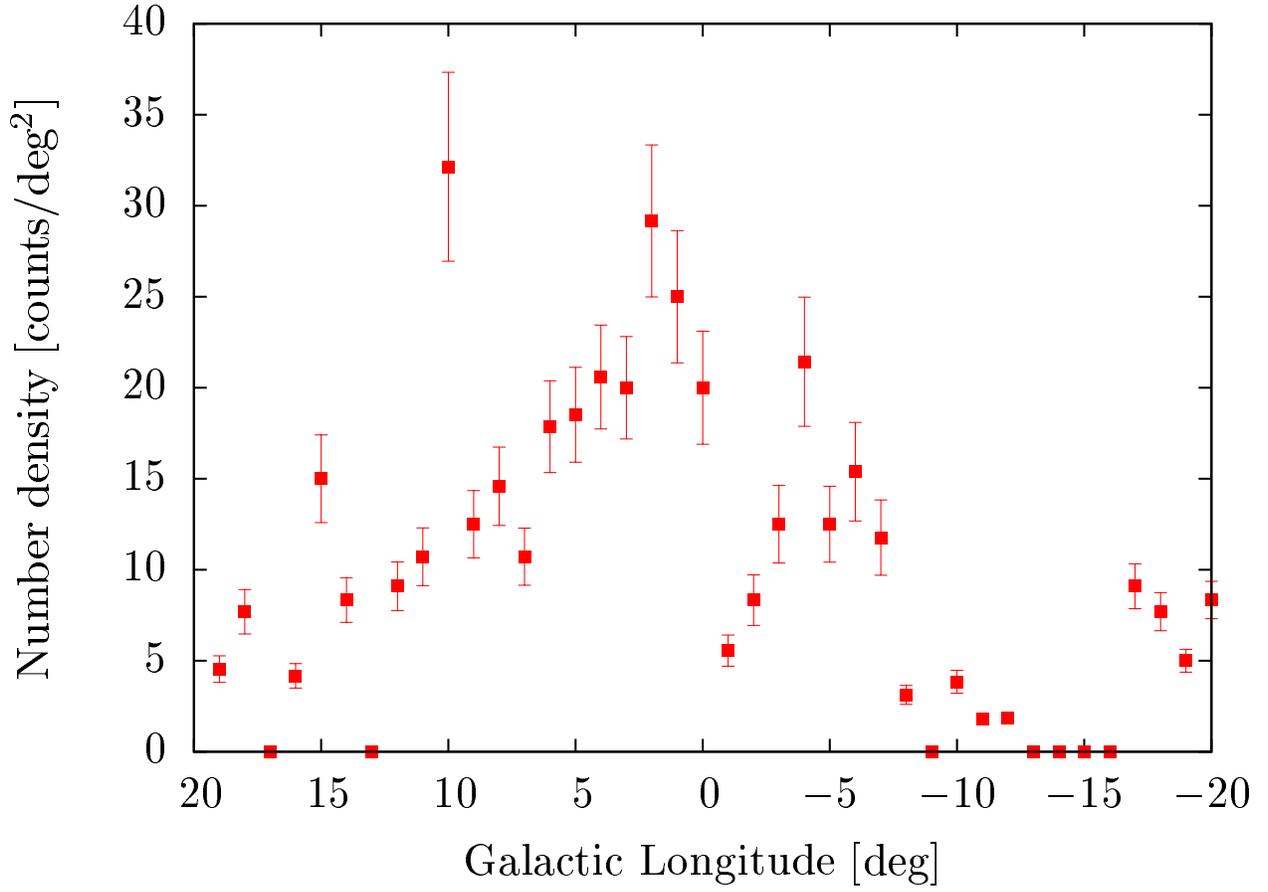}}
		\caption{Number density distribution for the detected objects every $l=1^\circ$.
		\label{Number-Density}}
	\end{center}
\end{figure}

\begin{figure}[thbp]
	\begin{center}
		\resizebox{\hsize}{!}{\includegraphics[width=6cm,clip]{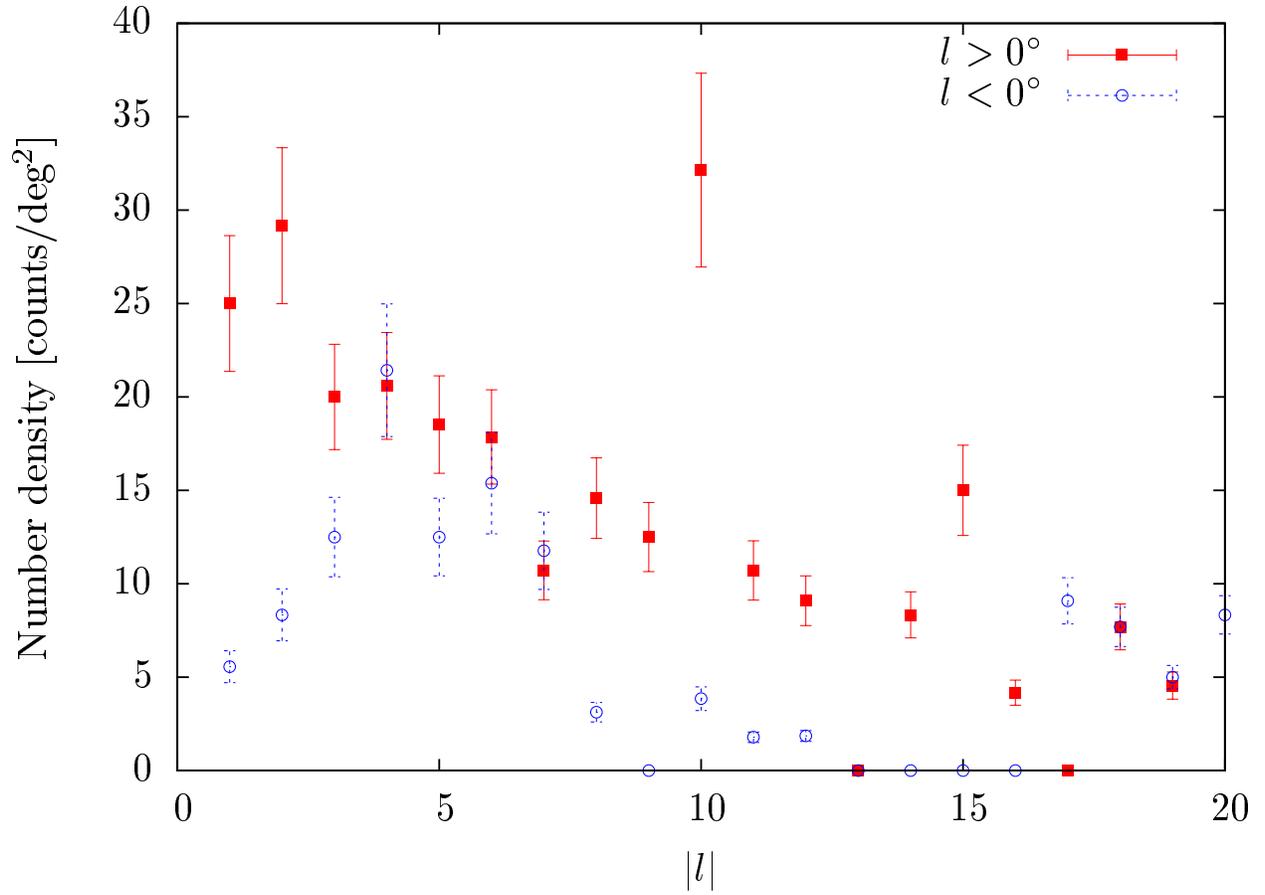}}
		\caption{The comparison of the number densities between $l>0^\circ$ and $l<0^\circ$. 
					The red squares represent the number densities in $l>0^\circ$, and 
					the blue circles represent the number densities in $l<0^\circ$, respectively. 
		\label{Number-Density-absolutelon}}
	\end{center}
\end{figure}

Figure \ref{Number-Density} shows the number density distribution for the detected variable stars. 
It shows an asymmetric distribution towards $l=0^\circ$, and 
the number density in $l>0^\circ$ is still relatively larger than that in $l<0^\circ$. 
One maybe consider that the asymmetry is caused by a small number of samples. 
However, if the Galactic bulge has a symmetric structure, 
it is strange that in most of the $|l|$ bins (see Fig. \ref{Number-Density-absolutelon}) 
the number densities in $l>0^\circ$ are larger than that in $l<0^\circ$ 
(especially, the differences in $l\sim2^\circ$ and $l\sim10^\circ$ are clear). 
In other words, 
if a small number statistics cause the asymmetry, 
the number densities in $l<0^\circ$ should evenly be 
nearly or larger than that in $l>0^\circ$. 
In addition, the statistical test also support the asymmetry. 
We performed $\chi^2$ test to examine the significance of this asymmetry, 
assuming the null hypothesis that the number density is a symmetric distribution 
towards $l=0^\circ$ within $|l| \leq 10^\circ$. 
We derived $\chi^2=35.4$, 
which implies that the difference of the distributions between $l>0^\circ$ and $l<0^\circ$ 
is significant at the 99$\%$ confidence level. 
Therefore, it is highly probable that the number density distribution is asymmetry, 
which is not due to a small number of samples. 
Below, we consider possible causes for this asymmetry. 

First, the difference of the number of investigated images might affect the number of the detection. 
However, the number of investigated images in each region is 
619 ($0^\circ \leq l \leq 10^\circ$, $|b| \leq 5^\circ$) and 
461 ($-10^\circ \leq l \leq 0^\circ$, $|b| \leq 5^\circ$) images (see Table \ref{detailed-bulge}), 
which correspond to the average number densities of 17.4 and 7.1 counts deg$^{-2}$, respectively. 
Therefore, the difference between the numbers of investigated images can not explain the asymmetry. 

As demonstrated in Sect. \ref{Probability}, the difference among epoch interval in the overlapping regions 
less affect the detection probability. 
Accordingly, it can not be the cause for the asymmetry. 

Many variable stars are expected to be detected within or near the Galactic disk ($|b| \lesssim 1^\circ$), 
so if the number of investigated images is different between $l>0^\circ$ and $l<0^\circ$, 
the difference of number density might be significant. 
However, they are 75 ($0^\circ \leq l \leq 10^\circ$, $|b| \leq 1^\circ$) and 
98 ($-10^\circ \leq l \leq 0^\circ$, $|b| \leq 1^\circ$) images, respectively. 
Hence, the asymmetry can not be explained 
by the difference of the number of the investigated images in $|b| \leq 1^\circ$. 

The local heavy interstellar extinction might cause the difference of the number densities 
between $l>0^\circ$ and $l<0^\circ$ 
because we may miss variable stars in such region, 
although we can observe the deep bulge at J-band to some extent. 
However, as mentioned in the above, in the most of the $|l|$ bins the number densities in $l>0^\circ$ 
are larger than those in $l<0^\circ$. 
Accordingly, it is difficult to explain these differences by the local highly extincted regions. 
Still, we note that there are drop-offs at $l=-1^\circ, -2^\circ$. 
This is strange because the number density around $l\sim 0^\circ$ 
is expected to be larger than that in $l<-2^\circ$. 
According to the extinction map of \citet{Dutra2003-MNRAS}, 
the extinction around $l\sim -1^\circ$ is slightly larger than other region. 
Therefore, the drop-offs might be caused by the relatively larger extinction. 

High-mass losing stars such as OH/IR stars may be missed due to heavy circumstellar extinction. 
However, such highly obscured stars should be uniformly undetected at J-band. 
The OH/IR candidates detected in this paper are considered to be relatively low circumstellar extinction. 
If objects are OH/IR stars having low circumstellar dust enough to be detected at J-band, 
they can be detected at constant detection probability as demonstrated in Sect. \ref{Detection-probability}.

Finally, it is considered that the asymmetry reflect the structure of the Galactic bulge. 
Various studies reveal that the Galactic bulge has an asymmetric structure, that is, 
there is a central bar in the bulge \citep{Nakada1991-Nature,Blitz1991-ApJ,Stanek1994-ApJL}. 
The bar tilt with respect to the line toward the Galactic centre, 
and the closer end lie in the positive Galactic longitude \citep[e.g., ][]{Morris1996-ARAA}. 
Therefore, the asymmetry of the number density distribution 
can be explained by the tilted bar structure in the Galactic bulge. 
As mentioned by \citet{Whitelock1992-ASSL} and \citet{Matsunaga2005-MNRAS}, 
Mira variables can trace the structure of the Galactic bulge. 
Hence, it is natural consequence that 
our large-amplitude variable stars also trace the structure of the bulge.

\section{Summary and Conclusions}

We have discovered variable stars using overlapping regions in the 2MASS public images. 
As a result of the investigation toward the Galactic centre 
($-30^\circ \lesssim l \lesssim 20^\circ, |b| \lesssim 20^\circ$), 
we have detected 136 variable stars. 
Among which, 6 variable stars are already-known and 
84 are accompanied by an optical counterpart in DSS images. 
The optical counterparts, however, are too faint to measure magnitudes
(i.e., their brightness are nearby limiting magnitude of DSS images).
The interstellar extinctions are estimated on the basis of the position of the upper giant branch, 
which are consistent with values of \citet{Dutra2003-MNRAS}. 
Photometric properties of the detected variable stars indicate that 
most of them are large-amplitude AGB variables such as Mira variables and OH/IR stars in the Galactic bulge. 
Additionally, the bar-like structure of the bulge is detected 
by the number density distribution of the detected variable stars. 

This paper demonstrates that the search for variable stars using overlapping regions is an useful method. 
In this paper, there is a small number of detection due to the strict detection criteria, 
but it is possible to detect a lot of variable stars 
if we can judge the variability more precisely with tender criteria. 
Establishing this search method, we can discover variable stars in widely region. 
If extracting more variable stars in the bulge, the structure of the bulge might be revealed using them as tracers, 
since the detected variable stars in this paper trace the structure of the bulge.

\acknowledgments

This publication makes use of data products from the Two Micron All Sky Survey, 
which is a joint project of the University of Massachusetts and 
the Infrared Processing and Analysis Center/California Institute of Technology, 
funded by the National Aeronautics and Space Administration and the National Science Foundation.
TOPCAT (http://www.starlink.ac.uk/topcat/) helps us investigate the photometric properties of the detected variables. 
S.K. is grateful to Noriyuki Matsunaga for valuable comments and discussions. 
We thank the anonymous referee for many comments to improve this paper.

\clearpage

\clearpage

\begin{deluxetable}{rccrrrllrrrc}
\tabletypesize{\scriptsize}
\tablecaption{Catalogue of the detected variable stars. 
The $\Delta m_\textnormal{\tiny J}$ represents the light variance at J-band between two epoch measurements, 
and the $\Delta d$ represents the interval of epochs. 
The O.C. means optical counterpart. 
\label{catalogue}}
\tablewidth{0pt}
\tablehead{
\colhead{No.} & \colhead{RA} & \colhead{Dec} & \colhead{$J$} & \colhead{$H$} & \colhead{K} & 
\colhead{$A_\textnormal{\tiny K$_\textnormal{\tiny S}$}$} & \colhead{$\Delta m_\textnormal{J}$} & 
\colhead{$\Delta d$} & \colhead{$l$} & \colhead{$b$} & 
\colhead{O.C.} \\
\colhead{} & \colhead{[h m s]} & \colhead{[d m s]} & \colhead{[mag]} & \colhead{[mag]} & \colhead{[mag]} & 
\colhead{[mag]} & \colhead{[mag]} & \colhead{[d]} & \colhead{[deg]} & \colhead{[deg]} & \colhead{}
}
\startdata
$  1$ & 	 $17$ $41$ $17.1$   &	 $-25$ $12$ $35.0$   &	 $10.065$  &	 $8.661$   &	 $7.879$  &	 0.4197 &    1.18  &  368  &		 $2.667845$  &	 $2.775828$	&  $\bigcirc$  \\ 
$  2$ & 	 $17$ $41$ $16.7$   &	 $-28$ $44$ $28.7$   &	 $11.200$  &	 $8.963$   &	 $7.626$  &	 0.953  &    0.68  &  368  &		 $-0.333264$  &	 $0.911546$	&  $\times$  \\ $  3$ & 	 $17$ $41$ $19.9$   &	 $-27$ $54$ $31.0$   &	 $9.839$   &	 $8.481$   &	 $7.795$  &	 0.627  &    0.69  &  368  &		 $0.379878$  &	 $1.341830$ 	&  $\bigcirc$  \\ 
$  4$ & 	 $17$ $45$ $31.8$   &	 $-24$ $47$ $49.2$   &	 $10.081$  &	 $8.713$   &	 $8.055$  &	 0.5033 &    0.67  &  829  &		 $3.523988$  &	 $2.172698$ &  $\bigcirc$    \\ 
$  5$ & 	 $17$ $45$ $29.6$   &	 $-27$ $45$ $48.8$   &	 $12.171$  &	 $9.695$   &	 $8.367$  &	 1.3948 &    0.76  &  829  &		 $0.986317$  &	 $0.634843$ &  $\times$    \\ $  6$ & 	 $17$ $47$ $08.5$   &	 $-24$ $46$ $44.8$   &	 $10.057$  &	 $8.138$   &	 $6.954$  &	 0.5459 &    0.72  &  815  &		 $3.729208$  &	 $1.869280$ &  $\times$    \\ 
$  7$ & 	 $17$ $59$ $52.1$   &	 $-25$ $23$ $53.2$   &	 $10.208$  &	 $7.989$   &	 $6.600$  &	 0.9952 &    1.19  &  354  &		 $4.657897$  &	 $-0.927749$&  $\bigcirc$    \\ 
$  8$ & 	 $17$ $42$ $51.7$   &	 $-27$ $56$ $57.1$   &	 $10.073$  &	 $8.412$   &	 $7.569$  &	 0.7952 &    0.79  &  368  &		 $0.523589$  &	 $1.033264$ &  $\times$    \\ 
$  9$ & 	 $17$ $45$ $33.3$   &	 $-25$ $10$ $45.7$   &	 $8.106$   &	 $6.582$   &	 $5.655$  &	 0.4723 &    0.64  &  828  &		 $3.200183$  &	 $1.968936$ 	&  $\bigcirc$  \\
$ 10$ & 	 $17$ $45$ $31.5$   &	 $-26$ $41$ $42.9$   &	 $10.318$  &	 $9.051$   &	 $8.129$  &	 0.5839 &    0.94  &  829  &		 $1.901933$  &	 $1.185473$ &  $\bigcirc$	    \\ 
$ 11$ & 	 $18$ $02$ $29.6$   &	 $-26$ $50$ $33.9$   &	 $9.449$   &	 $8.224$   &	 $7.521$  &	 0.3794 &    0.65  &  473  &		 $3.693683$  &	 $-2.154590$ 	&  $\bigcirc$  \\ 
$ 12$ & 	 $18$ $02$ $29.7$   &	 $-28$ $14$ $09.8$   &	 $8.520$   &	 $7.530$   &	 $7.008$  &	 0.1302 &    0.84  &  473  &		 $2.479689$  &	 $-2.840399$ 	&  $\bigcirc$  \\ $ 13$ & 	 $18$ $02$ $35.6$   &	 $-28$ $50$ $53.4$   &	 $9.549$   &	 $7.917$   &	 $6.767$  &	 0.1345 &    1.01  &  473  &		 $1.956427$  &	 $-3.159922$ 	&  $\bigcirc$  \\ 
$ 14$ & 	 $18$ $02$ $34.5$   &	 $-29$ $15$ $04.3$   &	 $12.710$  &     $10.028$  &     $8.077$  &	 0.1614 &    1.52  &  473  &		 $1.602750$  &	 $-3.354251$ 	&  $\bigcirc$  \\ 
$ 15$ & 	 $18$ $06$ $15.0$   &	 $-24$ $29$ $46.7$   &	 $11.636$  &     $9.967$   &	 $8.830$  &	 0.6454 &    1.29  &  108  &		 $6.155342$  &	 $-1.742981$ 	&  $\bigcirc$  \\ 
$ 16$ & 	 $18$ $06$ $16.7$   &	 $-25$ $07$ $39.9$   &	 $10.489$  &     $9.143$   &	 $8.247$  &	 0.5357 &    0.68  &  108  &		 $5.606773$  &	 $-2.056296$ 	&  $\bigcirc$  \\ 
$ 17$ & 	 $18$ $06$ $17.0$   &	 $-26$ $13$ $45.6$   &	 $10.328$  &     $8.270$   &	 $6.969$  &	 0.3332 &    1.68  &  108  &		 $4.644303$  &	 $-2.593577$ 	&  $\times$  \\ 
$ 18$ & 	 $18$ $06$ $15.1$   &	 $-27$ $34$ $41.8$   &	 $8.475$   &	 $7.390$   &	 $6.721$  &	 0.1233 &    0.82  &  108  &		 $3.460492$  &	 $-3.243146$ 	&  $\bigcirc$  \\ 
$ 19$ & 	 $18$ $06$ $17.2$   &	 $-28$ $35$ $06.2$   &	 $9.410$   &	 $8.346$   &	 $7.821$  &	 0.1305 &    0.76  &  108  &		 $2.582295$  &	 $-3.738419$ 	&  $\bigcirc$  \\ 
$ 20$ & 	 $18$ $06$ $16.4$   &	 $-28$ $49$ $50.7$   &	 $8.813$   &	 $7.279$   &	 $6.216$  &	 0.1281 &    1.10  &  108  &		 $2.365496$  &	 $-3.854947$ 	&  $\times$  \\ 
$ 21$ & 	 $18$ $06$ $15.1$   &	 $-28$ $49$ $19.5$   &	 $9.367$   &	 $8.301$   &	 $7.489$  &	 0.128  &    0.61  &  108  &		 $2.370790$  &	 $-3.846596$ 	&  $\bigcirc$  \\ 
$ 22$ & 	 $17$ $45$ $30.2$   &	 $-25$ $28$ $18.1$   &	 $11.611$  &     $10.143$  &     $9.415$  &	 0.5318 &    0.59  &  829  &		 $2.944341$  & $1.826768$ 		&  $\times$  \\ 
$ 23$ & 	 $17$ $47$ $06.5$   &	 $-27$ $32$ $59.7$   &	 $12.050$  &	 $9.382$   &	 $7.792$  &	 1.6414 &    0.67  &  814  &		 $1.354721$  &	 $0.440360$ &  $\times$	   \\
$ 24$ & 	 $18$ $14$ $18.0$   &	 $-27$ $31$ $01.7$   &	 $9.890$   &	 $8.575$   &	 $7.896$  &	 0.0644 &    0.78  &  489  &		 $4.370932$  &	 $-4.779833$ 	&  $\bigcirc$  \\ 
$ 25$ & 	 $18$ $14$ $14.2$   &	 $-27$ $44$ $07.3$   &	 $10.123$  &	 $8.299$   &	 $7.147$  &	 0.0603 &    0.88  &  489  &		 $4.171250$  & $-4.870537$ 	&  $\bigcirc$	   \\ 
$ 26$ & 	 $18$ $15$ $15.3$   &	 $-24$ $10$ $50.9$   &	 $9.745$   &	 $8.075$   &	 $6.967$  &	 0.2256 &    0.92  &  57   &		 $7.419364$  &	 $-3.389888$ 	&  $\bigcirc$  \\ 
$ 27$ & 	 $18$ $15$ $19.4$   &	 $-24$ $25$ $23.3$   &	 $10.141$  &	 $8.406$   &	 $7.254$  &	 0.223  &    0.46  &  58   &		 $7.212985$  & $-3.518457$ 	&  $\times$      \\ 
$ 28$ & 	 $18$ $17$ $23.2$   &	 $-24$ $51$ $43.7$   &	 $11.427$  &	 $8.946$   &	 $7.286$  &	 0.1379 &    0.85  &  409  &		 $7.046953$  & $-4.139021$ 	&  $\times$	    \\ 
$ 29$ & 	 $18$ $17$ $26.6$   &	 $-25$ $30$ $28.9$   &	 $9.871$   &	 $8.816$   &	 $8.087$  &	 0.1256 &    0.98  &  410  &		 $6.481274$  &	 $-4.453887$ 	&  $\bigcirc$  \\ 
$ 30$ & 	 $18$ $22$ $12.1$   &	 $-28$ $42$ $18.7$   &	 $10.498$  &	 $8.809$   &	 $7.614$  &	 0.1296 &    0.63  &  717  &		 $4.128261$  & $-6.874790$ 	&  $\bigcirc$	   \\ 
$ 31$ & 	 $18$ $22$ $43.3$   &	 $-27$ $27$ $29.1$   &	 $9.676$   &	 $8.162$   &	 $7.120$  &	 0.1541 &    0.86  &  717  &		 $5.297151$  &	 $-6.405309$ 	&  $\bigcirc$    \\ 
$ 32$ & 	 $17$ $30$ $46.8$   &	 $-38$ $44$ $07.5$   &	 $10.475$  &	 $8.437$   &	 $7.184$  &	 0.3534 &    1.25  &  345  &		 $-9.928858$  & $-2.637008$ &  $\times$	    \\ 
$ 33$ & 	 $17$ $33$ $07.6$   &	 $-39$ $54$ $35.7$   &	 $10.238$  &	 $9.447$   &	 $9.187$  &	 0.2064 &    0.60  &  18   &		 $-10.667076$  & $-3.657835$&  $\bigcirc$	    \\ 
$ 34$ & 	 $17$ $40$ $19.7$   &	 $-36$ $45$ $00.9$   &	 $10.473$  &	 $9.047$   &	 $8.056$  &	 0.377  &    0.96  &  82   &		 $-7.236433$  &	 $-3.158788$&  $\bigcirc$	    \\ 
$ 35$ & 	 $17$ $46$ $20.5$   &	 $-37$ $08$ $45.0$   &	 $10.064$  &	 $8.785$   &	 $7.872$  &	 0.1682 &    0.87  &  342  &		 $-6.945796$  &	 $-4.389961$&  $\bigcirc$	    \\ 
$ 36$ & 	 $17$ $49$ $55.6$   &	 $-36$ $35$ $14.0$   &	 $12.160$  &	 $9.903$   &	 $8.260$  &	 0.1085 &    0.93  &  70   &		 $-6.096144$  &	 $-4.719827$&  $\bigcirc$	    \\ 
$ 37$ & 	 $17$ $56$ $30.2$   &	 $-37$ $21$ $52.8$   &	 $8.891$   &	 $7.889$   &	 $7.119$  &	 0.1105 &    0.88  &  341  &		 $-6.110810$  &	 $-6.246061$ 	&  $\bigcirc$   \\ 
$ 38$ & 	 $18$ $34$ $56.7$   &	 $-29$ $42$ $43.3$   &	 $9.483$   &	 $8.715$   &	 $8.177$  &	 0.0791 &    0.86  &  348  &		 $4.470819$  &	 $-9.812575$ 	&  $\bigcirc$    \\ 
$ 39$ & 	 $17$ $30$ $45.5$   &	 $-22$ $01$ $22.3$   &	 $10.453$  &	 $9.322$   &	 $8.516$  &	 0.3397 &    0.91  &  849  &		 $4.076803$  &	 $6.516994$ &  $\bigcirc$	    \\ 
$ 40$ & 	 $17$ $47$ $04.4$   &	 $-21$ $50$ $03.9$   &	 $9.422$   &	 $8.257$   &	 $7.606$  &	 0.3175 &    0.59  &  71   &		 $6.244762$  &	 $3.404609$ 	&  $\bigcirc$    \\ 
$ 41$ & 	 $17$ $47$ $03.0$   &	 $-22$ $17$ $02.7$   &	 $10.667$  &	 $8.976$   &	 $7.924$  &	 0.2692 &    1.08  &  71   &		 $5.856182$  &	 $3.177161$ &  $\bigcirc$	    \\ 
$ 42$ & 	 $17$ $47$ $05.5$   &	 $-22$ $35$ $04.1$   &	 $10.061$  &	 $9.083$   &	 $8.621$  &	 0.2349 &    0.58  &  71   &		 $5.603542$  &	 $3.013805$ &  $\bigcirc$	     \\ 
$ 43$ & 	 $17$ $47$ $04.9$   &	 $-23$ $38$ $52.3$   &	 $10.187$  &	 $8.853$   &	 $7.868$  &	 0.2891 &    0.59  &  71   &		 $4.690922$  &	 $2.466233$ &  $\bigcirc$	      \\ 
$ 44$ & 	 $17$ $47$ $02.3$   &	 $-23$ $36$ $10.2$   &	 $12.613$  &	 $10.277$  &     $8.693$  &	 0.2856 &    1.26  &  71   &		 $4.724360$  &	 $2.498013$ &  $\times$	     \\ 
$ 45$ & 	 $17$ $48$ $33.9$   &	 $-21$ $48$ $46.1$   &	 $10.464$  &	 $8.693$   &	 $7.424$  &	 0.3339 &    1.24  &  84   &		 $6.441830$  &	 $3.118943$ &  $\times$	  \\ 
$ 46$ & 	 $17$ $48$ $32.0$   &	 $-22$ $12$ $35.3$   &	 $10.344$  &	 $9.207$   &	 $8.534$  &	 0.293  &    0.57  &  84   &		 $6.097000$  &	 $2.921243$ &  $\bigcirc$	  \\ 
$ 47$ & 	 $17$ $49$ $03.7$   &	 $-22$ $27$ $10.9$   &	 $9.745$   &	 $8.577$   &	 $7.829$  &	 0.3481 &    0.79  &  865  &		 $5.950905$  &	 $2.691479$ 	&  $\bigcirc$   \\ 
$ 48$ & 	 $17$ $51$ $04.2$   &	 $-23$ $18$ $27.7$   &	 $10.715$  &	 $8.986$   &	 $7.947$  &	 0.3759 &    1.29  &  100  &		 $5.453260$  &	 $1.856510$ &  $\bigcirc$	  \\ 
$ 49$ & 	 $17$ $51$ $03.2$   &	 $-23$ $17$ $34.0$   &	 $12.847$  &	 $10.079$  &     $8.183$  &	 0.3754 &    1.25  &  100  &		 $5.464136$  &	 $1.867419$ &  $\times$	  \\ 
$ 50$ & 	 $17$ $52$ $05.0$   &	 $-20$ $38$ $21.3$   &	 $11.088$  &	 $9.878$   &	 $9.202$  &	 0.3483 &    0.81  &  867  &		 $7.871569$  &	 $3.013364$ &  $\bigcirc$	  \\ 
$ 51$ & 	 $17$ $52$ $04.4$   &	 $-22$ $16$ $09.4$   &	 $8.583$   &	 $7.172$   &	 $6.286$  &	 0.3056 &    0.96  &  867  &		 $6.465151$  &	 $2.186872$ 	&  $\bigcirc$   \\ 
$ 52$ & 	 $17$ $54$ $42.1$   &	 $-23$ $34$ $52.6$   &	 $11.087$  &	 $9.014$   &	 $7.750$  &	 0.7294 &    1.05  &  349  &		 $5.640794$  &	 $1.000049$ &  $\times$	  \\ 
$ 53$ & 	 $18$ $03$ $21.1$   &	 $-20$ $08$ $05.3$   &	 $12.455$  &	 $10.208$  &     $8.971$  &	 1.1573 &    0.70  &  322  &		 $9.629374$  &	 $0.978341$ &  $\bigcirc$	  \\ 
$ 54$ & 	 $18$ $03$ $22.1$   &	 $-20$ $00$ $40.7$   &	 $10.221$  &	 $7.646$   &	 $6.195$  &	 1.1314 &    0.99  &  322  &		 $9.738836$  &	 $1.035692$ &  $\times$	  \\ 
$ 55$ & 	 $18$ $03$ $21.2$   &	 $-20$ $00$ $15.7$   &	 $11.785$  &	 $9.502$   &	 $8.196$  &	 1.1251 &    0.77  &  322  &		 $9.743150$  &	 $1.042176$ &  $\times$	   \\ 
$ 56$ & 	 $18$ $03$ $21.3$   &	 $-20$ $45$ $57.0$   &	 $11.116$  &	 $8.506$   &	 $7.153$  &	 0.9228 &    2.37  &  322  &		 $9.080352$  &	 $0.667159$ &  $\times$	   \\ $ 57$ & 	 $18$ $07$ $23.9$   &	 $-19$ $06$ $20.2$   &	 $11.771$  &	 $9.126$   &	 $7.606$  &	 1.0491 &    1.26  &  332  &		 $10.993322$  &	 $0.650770$ &  $\times$	   \\ 
$ 58$ & 	 $18$ $07$ $23.8$   &	 $-19$ $29$ $50.0$   &	 $11.840$  &	 $9.628$   &	 $8.322$  &	 1.1183 &    1.01  &  332  &		 $10.650984$  &	 $0.460572$ &  $\times$	   \\ 
$ 59$ & 	 $18$ $07$ $26.9$   &	 $-19$ $28$ $40.0$   &	 $9.756$   &	 $7.387$   &	 $6.091$  &	 1.1136 &    1.08  &  332  &		 $10.673897$  &	 $0.459394$ &  $\bigcirc$	    \\ 
$ 60$ & 	 $18$ $07$ $25.9$   &	 $-21$ $25$ $28.4$   &	 $12.746$  &	 $9.400$   &	 $7.525$  &	 1.9955 &    0.86  &  332  &		 $8.971180$  &	 $-0.484417$&  $\times$	   \\ 
$ 61$ & 	 $18$ $07$ $26.6$   &	 $-21$ $52$ $06.6$   &	 $13.049$  &	 $9.952$   &	 $8.202$  &	 1.5003 &    0.96  &  332  &		 $8.584611$  &	 $-0.702762$&  $\times$	    \\ 
$ 62$ & 	 $18$ $08$ $57.2$   &	 $-22$ $38$ $56.4$   &	 $10.838$  &	 $8.933$   &	 $7.801$  &	 0.87   &    1.20  &  332  &		 $8.071647$  &	 $-1.387010$&  $\times$	    \\ 
$ 63$ & 	 $18$ $10$ $28.4$   &	 $-20$ $10$ $36.6$   &	 $13.543$  &	 $9.946$   &	 $7.842$  &	 1.4913 &    1.30  &  404  &		 $10.407129$  &	 $-0.501659$&  $\bigcirc$	   \\ 
$ 64$ & 	 $18$ $10$ $31.2$   &	 $-20$ $18$ $20.8$   &	 $11.508$  &	 $8.465$   &	 $6.379$  &	 1.4858 &    1.24  &  404  &		 $10.299469$  &	 $-0.573469$&  $\times$	   \\ 
$ 65$ & 	 $18$ $14$ $37.2$   &	 $-23$ $22$ $45.5$   &	 $10.319$  &	 $9.302$   &	 $8.527$  &	 0.2195 &    0.96  &  361  &		 $8.056689$  &	 $-2.881419$&  $\bigcirc$	    \\ 
$ 66$ & 	 $18$ $22$ $43.2$   &	 $-22$ $28$ $03.4$   &	 $10.177$  &	 $8.754$   &	 $7.922$  &	 0.1203 &    1.43  &  372  &		 $9.741503$  &	 $-4.098614$&  $\bigcirc$	    \\ 
$ 67$ & 	 $17$ $59$ $57.3$   &	 $-16$ $11$ $36.8$   &	 $8.599$   &	 $7.550$   &	 $7.009$  &	 0.1653 &    0.84  &  116  &		 $12.662381$  &	 $3.624369$ 	&  $\bigcirc$   \\ 
$ 68$ & 	 $18$ $23$ $47.1$   &	 $-14$ $48$ $52.3$   &	 $13.681$  &	 $10.510$  &     $8.690$  &	 1.5986 &    0.84  &  514  &		 $16.633148$  &	 $-0.746451$&  $\times$	    \\
$ 69$ & 	 $18$ $23$ $45.3$   &	 $-15$ $34$ $54.8$   &	 $13.004$  &	 $10.577$  &     $9.167$  &	 1.2218 &    0.82  &  514  &		 $15.951609$  &	 $-1.099328$&  $\times$	    \\ 
$ 70$ & 	 $18$ $23$ $44.4$   &	 $-15$ $40$ $17.9$   &	 $12.401$  &	 $10.352$  &     $9.146$  &	 0.9239 &    0.78  &  514  &		 $15.870594$  &	 $-1.138150$&  $\times$	    \\ 
$ 71$ & 	 $18$ $26$ $13.0$   &	 $-16$ $44$ $26.8$   &	 $11.369$  &	 $9.137$   &	 $7.649$  &	 0.4598 &    0.97  &  360  &		 $15.202174$  &	 $-2.162917$&  $\times$	    \\ 
$ 72$ & 	 $18$ $28$ $41.7$   &	 $-13$ $37$ $28.0$   &	 $11.475$  &	 $9.913$   &	 $8.954$  &	 0.7364 &    0.64  &  429  &		 $18.240893$  &	 $-1.244478$&  $\bigcirc$	    \\ 
$ 73$ & 	 $19$ $06$ $24.4$   &	 $-15$ $09$ $53.5$   &	 $9.739$   &	 $7.830$   &	 $6.576$  &	 0.0753 &    1.28  &  132  &		 $20.967398$  &	 $-10.098936$ 	&  $\bigcirc$   \\ 
$ 74$ & 	 $17$ $31$ $40.2$   &	 $-34$ $11$ $37.2$   &	 $11.380$  &	 $10.781$  &     $10.438$ &	 1.2133 &    0.55  &  4    &	  $-6.030693$  &	 $-0.296953$&  $\bigcirc$	    \\ 
$ 75$ & 	 $17$ $41$ $15.3$   &	 $-31$ $33$ $56.3$   &	 $13.864$  &	 $11.247$  &     $10.166$ &	 1.5635 &    0.68  &  325  &	  $-2.732807$  &	 $-0.578236$&  $\times$	    \\ 
$ 76$ & 	 $17$ $41$ $13.5$   &	 $-32$ $03$ $50.1$   &	 $12.045$  &	 $9.588$   &	 $8.199$  &	 1.0715 &    1.43  &  325  &		 $-3.159065$  &	 $-0.836410$&  $\times$	    \\ 
$ 77$ & 	 $17$ $41$ $13.3$   &	 $-32$ $36$ $44.8$   &	 $10.256$  &	 $7.878$   &	 $6.399$  &	 0.7088 &    1.57  &  325  &		 $-3.625026$  &	 $-1.125940$&  $\times$	    \\ 
$ 78$ & 	 $17$ $41$ $17.3$   &	 $-34$ $02$ $14.7$   &	 $10.453$  &	 $8.682$   &	 $7.575$  &	 0.4342 &    0.96  &  325  &		 $-4.827726$  &	 $-1.890973$&  $\bigcirc$	    \\ 
$ 79$ & 	 $17$ $41$ $12.4$   &	 $-34$ $48$ $24.1$   &	 $8.261$   &	 $6.805$   &	 $5.823$  &	 0.3615 &    1.19  &  325  &		 $-5.490286$  &	 $-2.282981$ 	&  $\bigcirc$   \\ 
$ 80$ & 	 $17$ $41$ $11.8$   &	 $-34$ $47$ $00.0$   &	 $9.366$   &	 $8.129$   &	 $7.532$  &	 0.3618 &    0.72  &  325  &		 $-5.471517$  &	 $-2.268901$ 	&  $\bigcirc$   \\ 
$ 81$ & 	 $17$ $42$ $56.3$   &	 $-32$ $04$ $17.9$   &	 $13.374$  &	 $10.862$  &     $9.528$  &	 1.2253 &    0.85  &  325  &		 $-2.974196$  &	 $-1.148894$&  $\times$	    \\ 
$ 82$ & 	 $17$ $42$ $56.6$   &	 $-32$ $24$ $52.7$   &	 $11.106$  &	 $9.319$   &	 $8.417$  &	 0.9779 &    0.64  &  325  &		 $-3.265510$  &	 $-1.330074$&  $\times$	    \\ 
$ 83$ & 	 $17$ $42$ $52.6$   &	 $-33$ $43$ $39.5$   &	 $8.625$   &	 $7.145$   &	 $6.238$  &	 0.4884 &    1.57  &  325  &		 $-4.390516$  &	 $-2.008022$ 	&  $\bigcirc$   \\ 
$ 84$ & 	 $17$ $42$ $54.5$   &	 $-35$ $04$ $18.7$   &	 $10.302$  &	 $9.228$   &	 $8.587$  &	 0.268  &    0.91  &  325  &		 $-5.532279$  &	 $-2.719068$&  $\bigcirc$	    \\ 
$ 85$ & 	 $19$ $04$ $34.0$   &	 $-22$ $31$ $25.7$   &	 $14.626$  &     $13.967$  &     $13.939$ &	 0.065  &    1.70  &  409  &	  $13.972520$  &	 $-12.811301$ 	&  $\bigcirc$   \\ 
$ 86$ & 	 $17$ $12$ $36.3$   &	 $-28$ $06$ $57.0$   &	 $10.666$  &	 $8.809$   &	 $7.604$  &	 0.2221 &    1.21  &  725  &		 $-3.319054$  &	 $6.501491$ &  $\times$	    \\ 
$ 87$ & 	 $17$ $12$ $37.3$   &	 $-28$ $10$ $59.6$   &	 $8.387$   &	 $7.366$   &	 $6.751$  &	 0.2212 &    0.90  &  726  &		 $-3.372093$  &	 $6.459331$ 	&  $\bigcirc$   \\ 
$ 88$ & 	 $17$ $16$ $56.7$   &	 $-22$ $28$ $59.3$   &	 $11.848$  &	 $9.345$   &	 $7.536$  &	 0.2586 &    0.87  &  110  &		 $1.906583$  &	 $8.925828$ &  $\times$	    \\ 
$ 89$ & 	 $16$ $58$ $04.9$   &	 $-46$ $02$ $36.4$   &	 $10.063$  &	 $7.890$   &	 $6.491$  &	 0.9234 &    1.16  &  81   &		 $-19.384462$  & $-2.006203$&  $\bigcirc$	    \\ 
$ 90$ & 	 $17$ $08$ $28.9$   &	 $-43$ $53$ $32.6$   &	 $9.048$   &	 $7.395$   &	 $6.409$  &	 0.5734 &    1.31  &  205  &		 $-16.556521$  &	 $-2.155356$&  $\bigcirc$   \\ 
$ 91$ & 	 $17$ $08$ $30.8$   &	 $-45$ $40$ $01.5$   &	 $11.812$  &	 $9.612$   &	 $8.127$  &	 0.3058 &    1.35  &  205  &		 $-17.979133$  & $-3.218489$&  $\bigcirc$	    \\ 
$ 92$ & 	 $16$ $30$ $37.5$   &	 $-26$ $43$ $33.1$   &	 $16.655$  &	 $16.000$  &     $15.613$ &	 0.1324 &    ---   &  369  &	  $-8.095246$  &	 $14.665882$&  $\bigcirc$	    \\ 
$ 93$ & 	 $16$ $30$ $37.4$   &	 $-26$ $44$ $54.2$   &	 $9.997$   &	 $7.751$   &	 $6.107$  &	 0.1321 &    1.16  &  369  &		 $-8.113004$  &	 $14.651304$ 	&  $\bigcirc$   \\ 
$ 94$ & 	 $18$ $44$ $31.0$   &	 $-30$ $37$ $09.8$   &	 $13.236$  &	 $11.183$  &     $9.612$  &	 0.0457 &    1.11  &  143  &		 $4.529211$  & $-12.070860$ 	&  $\bigcirc$	\\ 
$ 95$ & 	 $18$ $44$ $30.2$   &	 $-30$ $41$ $10.0$   &	 $9.540$   &	 $8.209$   &	 $7.240$  &	 0.0376 &    0.91  &  143  &		 $4.465919$  &	 $-12.095953$ 	&  $\bigcirc$   \\ 
$ 96$ & 	 $17$ $42$ $51.1$   &	 $-27$ $26$ $54.1$   &	 $9.883$   &	 $8.436$   &	 $7.577$  &	 0.6121 &    0.38  &  368  &		 $0.948530$  &	 $1.298480$ 	&  $\bigcirc$   \\ 
$ 97$ & 	 $18$ $02$ $33.4$   &	 $-28$ $53$ $56.2$   &	 $10.896$  &	 $8.298$   &	 $6.545$  &	 0.1376 &    0.53  &  473  &		 $1.908176$  &	 $-3.177860$&  $\times$	    \\ 
$ 98$ & 	 $18$ $06$ $16.9$   &	 $-24$ $56$ $15.0$   &	 $9.274$   &	 $7.814$   &	 $6.919$  &	 0.5213 &    0.47  &  108  &		 $5.773390$  &	 $-1.964271$ 	&  $\bigcirc$   \\ 
$ 99$ & 	 $18$ $11$ $31.9$   &	 $-28$ $47$ $56.1$   &	 $8.201$   &	 $7.057$   &	 $6.360$  &	 0.0756 &    0.54  &  489  &		 $2.948058$  &	 $-4.850066$ 	&  $\bigcirc$   \\ 
$100$ & 	 $17$ $47$ $07.1$   &	 $-24$ $36$ $21.1$   &	 $10.525$  &	 $9.206$   &	 $8.470$  &	 0.4949 &    0.38  &  814  &		 $3.874789$  &	 $1.963479$ &  $\bigcirc$	    \\ 
$101$ & 	 $18$ $14$ $13.7$   &	 $-25$ $24$ $10.5$   &	 $10.251$  &	 $9.154$   &	 $8.363$  &	 0.2488 &    0.43  &  489  &		 $6.230664$  &	 $-3.764513$&  $\bigcirc$	    \\ 
$102$ & 	 $18$ $14$ $12.7$   &	 $-27$ $43$ $28.6$   &	 $8.887$   &	 $8.044$   &	 $7.451$  &	 0.0599 &    0.40  &  489  &		 $4.178138$  &	 $-4.860584$ 	&  $\bigcirc$   \\ 
$103$ & 	 $18$ $15$ $19.6$   &	 $-25$ $38$ $22.2$   &	 $9.232$   &	 $8.010$   &	 $7.174$  &	 0.1484 &    0.54  &  58   &		 $6.139465$  &	 $-4.094807$ 	&  $\bigcirc$   \\ 
$104$ & 	 $18$ $17$ $22.7$   &	 $-25$ $17$ $18.3$   &	 $9.044$   &	 $8.156$   &	 $7.677$  &	 0.125  &    0.37  &  409  &		 $6.668801$  &	 $-4.337764$ 	&  $\bigcirc$   \\ 
$105$ & 	 $17$ $30$ $42.2$   &	 $-40$ $31$ $51.7$   &	 $8.931$   &	 $7.673$   &	 $6.954$  &	 0.159  &    0.53  &  344  &		 $-11.441486$  &	 $-3.608212$&  $\bigcirc$   \\ 
$106$ & 	 $17$ $40$ $22.4$   &	 $-39$ $45$ $54.0$   &	 $8.354$   &	 $7.344$   &	 $6.669$  &	 0.1446 &    0.59  &  81   &		 $-9.799220$  &	 $-4.756486$ 	&  $\bigcirc$   \\ 
$107$ & 	 $17$ $49$ $56.2$   &	 $-37$ $00$ $04.7$   &	 $9.371$   &	 $8.456$   &	 $7.974$  &	 0.1213 &    0.72  &  71   &		 $-6.452994$  &	 $-4.932027$ 	&  $\bigcirc$   \\ 
$108$ & 	 $17$ $30$ $57.4$   &	 $-22$ $10$ $41.1$   &	 $9.923$   &	 $8.829$   &	 $8.108$  &	 0.3877 &    0.64  &  850  &		 $3.970470$  &	 $6.394353$ 	&  $\bigcirc$   \\ 
$109$ & 	 $17$ $30$ $48.7$   &	 $-23$ $05$ $32.9$   &	 $9.582$   &	 $8.402$   &	 $7.680$  &	 0.339  &    0.64  &  849  &		 $3.179284$  &	 $5.926952$ 	&  $\bigcirc$   \\ 
$110$ & 	 $17$ $30$ $51.3$   &	 $-23$ $14$ $22.5$   &	 $9.242$   &	 $7.667$   &	 $6.628$  &	 0.3488 &    0.60  &  849  &		 $3.060500$  &	 $5.838750$ 	&  $\times$   \\ 
$111$ & 	 $17$ $48$ $32.9$   &	 $-21$ $17$ $56.0$   &	 $8.731$   &	 $7.479$   &	 $6.858$  &	 0.2638 &    0.63  &  84   &		 $6.881531$  &	 $3.386171$ 	&  $\bigcirc$   \\ 
$112$ & 	 $17$ $52$ $08.0$   &	 $-23$ $31$ $32.5$   &	 $9.372$   &	 $7.816$   &	 $6.951$  &	 0.4351 &    0.49  &  867  &		 $5.390104$  &	 $1.535504$ 	&  $\bigcirc$   \\ 
$113$ & 	 $17$ $54$ $10.0$   &	 $-23$ $06$ $36.1$   &	 $11.334$  &	 $9.039$   &	 $7.657$  &	 0.457  &    0.54  &  349  &		 $5.985368$  &	 $1.344251$ &  $\times$       \\ 
$114$ & 	 $17$ $54$ $42.0$   &	 $-17$ $51$ $50.5$   &	 $10.297$  &	 $9.212$   &	 $8.568$  &	 0.2087 &    0.63  &  346  &		 $10.583776$  &	 $3.881607$ &  $\bigcirc$       \\ 
$115$ & 	 $18$ $03$ $21.3$   &	 $-21$ $24$ $40.5$   &	 $12.439$  &	 $9.838$   &	 $8.239$  &	 1.4225 &    0.85  &  322  &		 $8.518488$  &	 $0.349514$ &  $\times$       \\ 
$116$ & 	 $18$ $03$ $19.5$   &	 $-22$ $48$ $30.7$   &	 $12.273$  &	 $9.908$   &	 $8.197$  &	 1.5636 &    0.61  &  322  &		 $7.298778$  &	 $-0.332218$&  $\times$       \\ 
$117$ & 	 $18$ $05$ $55.0$   &	 $-23$ $06$ $39.4$   &	 $11.377$  &	 $8.884$   &	 $7.322$  &	 1.1663 &    0.66  &  66   &		 $7.327730$  &	 $-1.000450$&  $\times$       \\ 
$118$ & 	 $18$ $07$ $27.1$   &	 $-19$ $46$ $39.5$   &	 $13.339$  &	 $9.356$   &	 $6.350$  &	 1.9458 &    0.78  &  332  &		 $10.412295$  &	 $0.312818$ &  $\times$       \\ 
$119$ & 	 $18$ $08$ $58.1$   &	 $-19$ $28$ $03.2$   &	 $13.463$  &	 $9.999$   &	 $8.097$  &	 1.4713 &    0.58  &  332  &		 $10.856788$  &	 $0.151157$ &  $\times$       \\ 
$120$ & 	 $18$ $08$ $59.4$   &	 $-22$ $00$ $03.0$   &	 $11.060$  &	 $8.716$   &	 $7.346$  &	 1.2381 &    0.60  &  332  &		 $8.643004$  &	 $-1.080586$&  $\times$       \\ $121$ & 	 $18$ $10$ $31.2$   &	 $-18$ $29$ $02.8$   &	 $11.563$  &	 $9.494$   &	 $8.277$  &	 0.9854 &    0.62  &  404  &		 $11.894999$  &	 $0.305629$ &  $\times$       \\ $122$ & 	 $18$ $10$ $29.8$   &	 $-19$ $01$ $36.8$   &	 $9.676$   &	 $7.439$   &	 $6.169$  &	 1.177  &    0.63  &  404  &		 $11.416946$  &	 $0.048520$ 	&  $\times$   \\ 
$123$ & 	 $18$ $10$ $30.5$   &	 $-22$ $31$ $30.4$   &	 $10.242$  &	 $8.610$   &	 $7.768$  &	 0.5533 &    0.58  &  404  &		 $8.353601$  &	 $-1.641445$&  $\times$       \\ 
$124$ & 	 $18$ $14$ $34.8$   &	 $-20$ $55$ $34.6$   &	 $10.320$  &	 $8.801$   &	 $7.934$  &	 0.5619 &    0.53  &  361  &		 $10.210830$  &	 $-1.704687$&  $\bigcirc$       \\ 
$125$ & 	 $18$ $18$ $40.2$   &	 $-18$ $29$ $26.3$   &	 $10.751$  &	 $9.971$   &	 $9.647$  &	 0.3935 &    0.52  &  16   &		 $12.812347$  &	 $-1.395134$&  $\bigcirc$       \\ $126$ & 	 $17$ $59$ $57.4$   &	 $-17$ $32$ $55.4$   &	 $10.021$  &	 $8.917$   &	 $8.299$  &	 0.2478 &    0.62  &  116  &		 $11.482901$  &	 $2.953623$ &@$\bigcirc$       \\ 
$127$ & 	 $18$ $23$ $46.7$   &	 $-12$ $49$ $56.5$   &	 $10.480$  &	 $7.898$   &	 $6.291$  &	 1.0722 &    0.52  &  514  &		 $18.383747$  &	 $0.183351$ &  $\times$       \\ 
$128$ & 	 $18$ $23$ $42.3$   &	 $-17$ $36$ $01.9$   &	 $11.045$  &	 $9.899$   &	 $9.293$  &	 0.2882 &    0.59  &  515  &		 $14.161179$  &	 $-2.033070$&  $\bigcirc$       \\ 
$129$ & 	 $18$ $26$ $10.0$   &	 $-17$ $43$ $34.9$   &	 $10.095$  &	 $8.999$   &	 $8.504$  &	 0.2063 &    0.67  &  360  &		 $14.323259$  &	 $-2.610543$&  $\times$       \\ 
$130$ & 	 $17$ $41$ $11.9$   &	 $-30$ $10$ $56.7$   &	 $13.088$  &	 $10.657$  &     $9.569$  &	 1.6505 &    0.78  &  325  &	 $-1.565476$ & $0.163940$ 	&  $\times$           \\ 
$131$ & 	 $17$ $42$ $55.2$   &	 $-35$ $14$ $49.5$   &	 $9.960$   &	 $8.858$   &	 $8.271$  &	 0.2417 &    0.70  &  325  &		 $-5.680403$  &	 $-2.812978$ 	&  $\bigcirc$   \\ 
$132$ & 	 $16$ $58$ $04.9$   &	 $-44$ $00$ $52.0$   &	 $10.211$  &	 $8.441$   &	 $7.519$  &	 0.8941 &    0.92  &  81   &	 $-17.794111$  & $-0.745574$ &  $\bigcirc$          \\ 
$133$ & 	 $16$ $58$ $06.6$   &	 $-45$ $31$ $46.1$   &	 $10.636$  &	 $9.144$   &	 $8.321$  &	 0.466  &    0.78  &  81   &		 $-18.978351$ & $-1.690865$ &  $\times$	    \\ 
$134$ & 	 $17$ $08$ $32.5$   &	 $-43$ $24$ $50.2$   &	 $11.835$  &	 $10.135$  &     $9.075$  &	 0.5238 &    0.64  &  205  &	 $-16.165976$  & $-1.878426$ &  $\bigcirc$          \\ 
$135$ & 	 $16$ $09$ $13.8$   &	 $-41$ $44$ $29.8$   &	 $6.921$   &	 $5.778$   &	 $5.021$  &  0.2805 &    0.73  &  92   & 	 $-22.202395$  &	 $7.362974$ &  $\bigcirc$ 	    \\ 
$136$ & 	 $18$ $28$ $32.3$   &	 $-11$ $44$ $11.6$   &	 $11.491$  &	 $8.586$   &	 $6.773$  &	 1.2861 &    1.01  &  432  &		 $19.895378$  &	 $-0.334072$&  $\times$	    \\ 
\enddata
\end{deluxetable}

\clearpage

\begin{table}
\caption{Detailed information for the bulge. 	
The $l$ values indicate the centre coordinates of each field. 
	For example, the $l=-17^\circ_{\cdot}5$ corresponds to 
    $-20^\circ \leq l \leq -15^\circ$. 
	The areas represent the investigated areas (total areas of the overlapping regions) in each fields 
	and provide approximate values. 
\label{detailed-bulge}}
\begin{center}\begin{tabular}{cc}
(a) $-5^\circ \leq b \leq 0^\circ $ & (b) $0^\circ \leq b \leq 5^\circ $  \\
\begin{tabular}{rrcc}
\hline \hline 
$l$ & $N_\textnormal{inv}$ & area & $N_\textnormal{var}$ \\ 
$[$deg$]$ & & [deg$^2$] &   \\ 
\hline
$-17.5$ & $99$  & $0.6875$ & $6$  \\
$-12.5$ & $251$ & $1.7419$ & $2$  \\
$-7.5$  & $263$ & $1.8147$ & $11$ \\
$-2.5$  & $68$  & $0.4692$ & $7$  \\
$2.5$   & $123$ & $0.8487$ & $15$ \\
$7.5$   & $216$ & $1.4904$ & $19$ \\
$12.5$  & $76$  & $0.5244$ & $6$  \\
$17.5$  & $125$ & $0.8625$ & $6$  \\ \hline
total   & $1221$& $8.4393$ & $72$ \\
\end{tabular}
&
\begin{tabular}{rrcc}
\hline \hline 
$l$ & $N_\textnormal{inv}$ & area & $N_\textnormal{var}$ \\
$[$deg$]$ & & [deg$^2$] &   \\ \hline
$-17.5$ &  $86$  & $0.5934$ & $0$  \\
$-12.5$ &  $97$  & $0.6693$ & $0$  \\
$-7.5$  &  $67$  & $0.4623$ & $0$  \\
$-2.5$  &  $63$  & $0.4347$ & $2$  \\
$2.5$   &  $83$  & $0.5727$ & $14$  \\
$7.5$   &  $197$ & $1.3593$ & $19$  \\
$12.5$  &  $110$ & $0.7590$ & $10$  \\
$17.5$  &  $43$  & $0.2967$ & $1$  \\ \hline
total   & $746$  & $5.1474$ & $46$ \\
\end{tabular}
\end{tabular}
\end{center}
\end{table}

\clearpage

\begin{deluxetable}{ll}
\tabletypesize{\scriptsize}
\tablecaption{List of the variable star catalogues which list up detected variables. 
		 The object numbers correspond to that in Table \ref{catalogue}. \label{VS-catalogues}}
\tablewidth{0pt}
\tablehead{
\colhead{Name of Catalogue} & \colhead{No. in the catalogue} 
}
\startdata
Combined General Catalogue of Variable Stars \citep{Samus2004-catalogue}     & No. 3, 33, 87  \\
Mira Variables in the OGLE Bulge fields \citep{Groenewegen2005-catalogue}    & No. 13, 97  \\
JKs photometry of Galactic bulge variables \citep{Schultheis2000-catalogue}  & No. 2  \\
Variables and Nebulae in Sgr B \citep{Terzan1995-catalogue}                  & No. 3, 87  \\
\enddata
\end{deluxetable}

\begin{deluxetable}{ll}
\tabletypesize{\scriptsize}
\tablecaption{List of the maser catalogues which list up detected variables. 
		 The object numbers correspond to that in Table \ref{catalogue}. \label{maser-catalogues}}
\tablewidth{0pt}
\tablehead{
\colhead{Name of Catalogue} & \colhead{No. in the catalogue} 
}
\startdata
86GHz SiO maser survey of late-type stars \citep{Messineo2002-AA}   & No. 23 \\
86GHz SiO maser survey in the Inner Galaxy \citep{Messineo2004-AA}     & No. 23  \\
1612MHz OH survey of IRAS point sources \citep{Lintel1991-AAS}    & No. 97, 118 \\
Galactic Bulge Region OH 1612MHz survey. I.  \citep{Sevenster1997-AAS}  & No. 97  \\
SiO masers in OH/IR stars, proto-PN and PN \citep{Nyman1998-AAS}        & No. 118  \\
\enddata
\end{deluxetable}

\end{document}